# Regular Rather than Chaotic Origin of the Resonant Transport in Superlattices

S. M. Soskin,[1,2,*] I. A. Khovanov,[3,4] and P. V. E. McClintock[2]

[1]*Institute of Semiconductor Physics, National Academy of Sciences of Ukraine, 03028 Kiev, Ukraine*
[2]*Physics Department, Lancaster University, Lancaster LA1 4YB, United Kingdom*
[3]*School of Engineering, University of Warwick, Coventry CV4 7AL, United Kingdom*
[4]*Centre for Scientific Computing, University of Warwick, Coventry CV4 7AL, United Kingdom*

We address the enhancement of electron transport in semiconductor superlattices that occurs in combined electric and magnetic fields when cyclotron rotation becomes resonant with Bloch oscillations. We show that the phenomenon is regular in origin, contrary to the widespread belief that it arises through chaotic diffusion. The theory verified by simulations provides an accurate description of earlier numerical results and suggests new ways of controlling resonant transport.

Spatial periodicity plays a fundamental role in nature. In particular, it governs quantum electron transport in crystals [1]. In a perfect crystal lattice, an electron in a constant electric field would undergo Bloch oscillations, moving forwards and backwards periodically so that its average drift would be zero [1]. But real lattices are imperfect, and electrons may be scattered before reversing their motion, allowing them to acquire a steady drift. Typically, the Bloch oscillation period $t_B$ greatly exceeds the average scattering time $t_s$, because $t_B$ is proportional to the reciprocal of the lattice period $d_l$, which is very small. So, Bloch oscillations are not observed in real crystals. Nanoscale superlattices [2] (SLs) impose on the crystal an additional periodicity with a period $d$ greatly exceeding $d_l$ but still small enough for the quantum nature of the electron to be important: $t_B$ may become comparable to or smaller than $t_s$ so that Bloch oscillations can manifest themselves, significantly suppressing the current, generating gigahertz or terahertz electric signals, and causing other important effects [3,4].

The first description of electron drift in SLs [2] showed that the drift velocity $v_d$ vs the electric field $F$ along a one-dimensional SL possesses a peak at $F = F_{\mathrm{ET}}$ such that $t_B$ (being $\propto F^{-1}$) is equal to $t_s$. It has important consequences, in particular, a peak in the differential conductivity vs voltage.

Another remarkable effect was predicted more recently [5,6]. It was noticed that, if a magnetic field is added, the dynamics reduces to that of an auxiliary classical harmonic oscillator at the cyclotron frequency subject to a traveling wave at the Bloch frequency. Numerical calculations within this model and the relaxation-time approximation for scattering [2] revealed additional peaks in $v_d(F)$ at the values of $F$ corresponding to integer ratios between the Bloch and cyclotron frequencies. As is known from the theory of dynamical systems, the phase plane of a harmonic oscillator subject to a traveling wave is threaded by a so-called stochastic web if the ratio between the wave and oscillator frequencies is an integer [7,8]. This web plays an important role in many physical systems [9,10]. It was conjectured [5,6] that the dynamical origin of the peaks lies in chaotic diffusion along the web. This conjecture stimulated wide interest and numerous theoretical and experimental investigations of the effect and its applications (e.g., Refs. [11–21]). These and many other works (e.g., Refs. [22–27]) assumed the original conjecture to be correct, implying that resonant electron transport in SLs can be controlled by chaotic diffusion [18].

In the present Letter, we show that this commonly held belief is incorrect: the peaks originate in a regular dynamics, while chaos, when present, destroys them.

Consider a one-dimensional SL. Because of the periodicity, it possesses minibands [2]. Let the SL parameters be such that only the lowest miniband is relevant [5,6,11,13–21]. The electron energy can [5,6] be approximated as $E(\vec{p}) = \Delta[1 - \cos(p_x d/\hbar)]/2 + (p_y^2 + p_z^2)/(2m^*)$, where $\vec{p} \equiv (p_x, p_y, p_z)$ is its quasimomentum, the $x$ axis is directed along the SL, $\Delta$ is the miniband width, $d$ is the SL period, and $m^*$ is the electron effective mass for motion in the transverse plane. Let us apply an electric field antiparallel to the SL axis and a magnetic field tilted at an angle $\theta < 90°$: $\vec{F} = (-F, 0, 0)$ and $\vec{B} = (B\cos(\theta), 0, B\sin(\theta))$, respectively. The semiclassical equations of motion are [1–5,14,17–19,28]

$$\frac{d\vec{p}}{dt} = -e\{\vec{F} + [\vec{v} \times \vec{B}]\},$$

$$\vec{v} \equiv \left(\frac{dx}{dt}, \frac{dy}{dt}, \frac{dz}{dt}\right) = \left(\frac{\partial E}{\partial p_x}, \frac{\partial E}{\partial p_y}, \frac{\partial E}{\partial p_z}\right), \quad (1)$$

where $e$ is the absolute value of the electronic charge.

The electron velocity in the $x$ direction is

$$v_x(t) \equiv \frac{dx}{dt} = \frac{\partial E}{\partial p_x} = \frac{\Delta d}{2\hbar}\sin\left(\frac{p_x(t)d}{\hbar}\right). \quad (2)$$

Within the relaxation-time approximation [2], with a correction allowing for the difference between the elastic and inelastic scattering, the drift velocity is [3,6,37]

$$v_d = \mu\nu\int_0^\infty dt\, e^{-\nu t} v_x(t), \quad \mu \equiv \sqrt{\frac{t_e}{t_e + t_i}}, \quad \nu \equiv \frac{1}{\mu t_i}, \quad (3)$$

where $t_e$ and $t_i$ are the elastic and inelastic scattering times, respectively.

If $B = 0$, then $p_x(t) = eFt$. So, $v_x(t) \propto \sin(\omega_B t)$ where $\omega_B \equiv edF/\hbar$ is the Bloch frequency, and Eq. (3) gives the modified Esaki-Tsu (ET) result [2]:

$$v_d(F) \equiv v_{\text{ET}}^{(\text{mod})}(\omega_B) = v_0^{(\text{mod})}\tilde{v}_{\text{ET}}\left(\frac{\omega_B}{\nu}\right),$$

$$\omega_B \equiv \frac{ed}{\hbar}F, \quad v_0^{(\text{mod})} \equiv \mu\frac{\Delta d}{2\hbar}, \quad \tilde{v}_{\text{ET}}(x) \equiv \frac{x}{1+x^2}. \quad (4)$$

The function $\tilde{v}_{\text{ET}}(\omega_B/\nu)$ (4) has a maximum at $\omega_B = \nu$.

If $B \neq 0$, the dynamics is much more complicated because the components of $\vec{p}$ are interwoven. Remarkably, however, the dynamics of $p_z$ reduces to a relatively simple form, and $p_x$ and $p_y$ can be expressed in terms of $p_z$ [5]. In terms of scaled quantities [19],

$$\frac{d^2\tilde{p}}{d\tilde{t}^2} + \tilde{p} = \epsilon \sin(\omega\tilde{t} - \tilde{p} + \phi_0),$$

$$\tilde{p} \equiv \tilde{p}_z(\tilde{t}) = p_z(t)\frac{d\tan(\theta)}{\hbar},$$

$$\tilde{t} \equiv \omega_c t, \quad \omega_c \equiv \frac{eB\cos(\theta)}{m^*}, \quad \omega \equiv \frac{\omega_B}{\omega_c},$$

$$\epsilon = \frac{\Delta m^*}{2}\left(\frac{d\tan(\theta)}{\hbar}\right)^2, \quad \phi_0 = p_{z0} + p_{x0},$$

$$p_{z0} \equiv \tilde{p}_z(0), \quad p_{x0} \equiv \tilde{p}_x(0), \quad \tilde{p}_x(\tilde{t}) = p_x(t)\frac{d}{\hbar}. \quad (5)$$

Two other scaled components of the momentum are related to $\tilde{p}_z(\tilde{t}) \equiv \tilde{p}(\tilde{t})$ as follows: $\tilde{p}_x(\tilde{t}) = p_{x0} + \omega\tilde{t} - (\tilde{p}_z(\tilde{t}) - p_{z0})$ and $\tilde{p}_y(\tilde{t}) \equiv p_y(t)d/\hbar = d\tilde{p}_z(\tilde{t})/d\tilde{t}$.

The physical origin of the dynamics (5), its relevance to $v_x(t)$, and the physical meanings of $\omega_c$ and $\epsilon$ are as follows. The transverse component of the magnetic field and electron motion along the SL generate a Lorentz force oscillating at frequency $\omega_B$. It excites a cyclotron rotation in the transverse plane which modulates $p_x$ and, via $p_x$, the angle of the Bloch oscillation. The frequency of the cyclotron rotation, which we will call the cyclotron frequency, is $\omega_c$. The amplitude of the Lorentz force in dimensionless units is $\epsilon$. For details, see Ref. [28].

We consider the case of zero temperature, which is the most important one [5,6,13–21]. Only zero initial momenta are then relevant [5,17,19,28]. So, the scaled drift velocity reads as

$$\tilde{v}_d \equiv \frac{v_d}{v_0^{(\text{mod})}} = \tilde{\nu}\int_0^\infty d\tilde{t}\, e^{-\tilde{\nu}\tilde{t}}\sin(\omega\tilde{t} - \tilde{p}), \quad \tilde{\nu} \equiv \frac{\nu}{\omega_c}, \quad (6)$$

where $\tilde{p} \equiv \tilde{p}(\tilde{t})$ is a solution of Eq. (5) with

$$\tilde{p}(0) = 0, \quad \frac{d\tilde{p}(\tilde{t}=0)}{d\tilde{t}} = 0, \quad \phi_0 = 0, \quad (7)$$

and $\tilde{\nu}$ is the scattering rate in terms of the dimensionless "time" $\tilde{t}$ (5).

We will show that the resonance peak in $\tilde{v}_d(\omega)$ at $\omega \approx 1$ may be of magnitude $\sim 1$ for arbitrarily small $\epsilon$. In contrast, the resonance contributions near multiple and rational frequencies necessarily vanish in the asymptotic limit $\epsilon \to 0$. These small contributions are ignored in our theory.

Necessary (but not sufficient) conditions for the distinct resonance peak are

$$\tilde{\nu} \ll 1, \quad \epsilon/4 \ll 1. \quad (8)$$

If any of these conditions fail, the resonant component of $v_x(t)$ cannot accumulate for long. Besides, if the second condition fails, the peaks at multiple or rational frequencies are significant and/or the dynamics at the relevant time scales is chaotic. We assume further that the conditions (8) hold true unless otherwise specified.

As is clear from Eqs. (5)–(7), the function $\tilde{v}_d(\omega)$ depends on two parameters: $\tilde{\nu}$ and $\epsilon$. But we show below that the magnitude and scaled shape of the resonance component depend only on a single parameter

$$\alpha \equiv \frac{\epsilon}{4\tilde{\nu}}. \quad (9)$$

It is proportional to the ratio of the two time scales—the scattering time and the time of the strong modulation of the Bloch oscillation angle—which in terms of dimensionless time (5) are $\tilde{t}_s = \tilde{\nu}^{-1}$ and $\tilde{t}_{\text{SM}} = \epsilon^{-1}$, respectively. To illustrate the latter time scale, consider the exact resonance $\omega_B = \omega_c$. The modulation amplitude $A_{\text{am}}$ then grows linearly with time, as $A_{\text{am}} = \epsilon\tilde{t}/2$, until $A_{\text{am}} \sim 1$. The latter range is reached just by $\tilde{t} \sim \tilde{t}_{\text{SM}}$, and so strong modulation essentially changes the dynamics (5). However, if $\alpha \ll 1$, then the scattering occurs before the modulation becomes strong, so that the latter is irrelevant. Otherwise, the strong modulation comes into play, and the drift enhancement occurs differently.

We consider first the limit $\alpha \ll 1$. In this case, the magnitude of $\tilde{p}$ at the scattering time scale $\tilde{t}_s \equiv \tilde{\nu}^{-1}$ is $\sim \alpha \ll 1$, so that we can neglect $\tilde{p}$ in $\sin(\omega\tilde{t} - \tilde{p})$ on the rhs of the equation of motion (5), which then reduces to the equation of the constrained vibration. Solving it with zero initial conditions (7) and substituting the result into the integrand of the integral (6), approximating $\sin(\omega \tilde{t} - \tilde{p})$ by $\sin(\omega \tilde{t}) - \cos(\omega \tilde{t})\tilde{p}$, integrating, and neglecting asymptotically small terms, we obtain

$$\tilde{v}_d = \tilde{v}_{\rm ET}(\omega/\tilde{\nu}) + \tilde{v}_{d,a}^{\rm (res)},$$
$$\tilde{v}_{d,a}^{\rm (res)} = \alpha \frac{2\omega/(1+\omega)}{1 + ((\omega-1)/\tilde{\nu})^2}, \qquad \alpha \ll 1. \qquad (10)$$

This is a superposition of the ET peak (4) and the resonance peak $\tilde{v}_{d,a}^{\rm (res)}(\omega)$. The latter has an asymptotically Lorentzian shape with a half-width $\tilde{\nu}$ and maximum $\alpha$ acquired at $\omega = 1$. The physical origin of the peak is as follows. If $\omega_B = \omega_c$, the modulation amplitude of the Bloch oscillation angle grows with time, while the modulation-induced deviation of $v_x(t)$ possesses a component that retains its sign and also grows with time, thus, being accumulated. If $\omega_B - \omega_c \neq 0$, the modulation amplitude grows more slowly, and, moreover, if $|\omega_B - \omega_c| \gg \nu$, the sign of the deviation changes many times during $t_s$ so that the drift averages to zero.

We now compare Eq. (10) with numerical simulations for the SL used in most experiments [6,11,13] and a typical magnetic field. So, let $d = 8.3$ nm, $\Delta = 19.1$ meV, $\nu = 4 \times 10^{12}$ s$^{-1}$, $m^* = 0.067 m_e$ (where $m_e$ is the free electron mass), and $B = 15$ T [13,19,20]. Then,

$$\tilde{\nu} \approx \frac{0.102}{\cos(\theta)}, \quad \epsilon \approx \frac{0.578}{\cot^2(\theta)}, \quad \alpha \approx 1.42 \frac{\sin^2(\theta)}{\cos(\theta)}. \quad (11)$$

Figure 1 presents the results for $\theta = 12°$ and $20°$, where $\alpha = 0.063$ and $0.177$, respectively. For $\theta = 12°$, the theory and simulations are virtually indistinguishable. For $\theta = 20°$, the theory only slightly exceeds the simulations.

As $\theta$ increases further, the excess of the theoretical resonant peak (10) over that in the simulations grows: $\tilde{v}_d(\omega = 1)$ in the simulations for $\theta = 40°$ [19,20] is about half that given by Eq. (10). The invalidity of Eq. (10) here is unsurprising because $\alpha \approx 0.77$ is not small.

To encompass arbitrary $\alpha$, we develop an approach suggested earlier [7,8] in a different context. If $\omega \simeq 1$

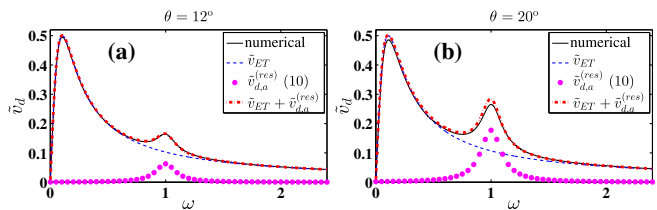

FIG. 1 (color online). Scaled drift velocity vs the ratio between the Bloch and cyclotron frequencies: comparison of numerical calculations (5)–(7) (black thin solid line) and the asymptotic theory (10) (red thick dash-dotted line) for (a) $\theta = 12°$, (b) $\theta = 20°$.

in Eq. (5), then, neglecting small fast oscillations, the dynamics reduces to that of the "resonant" Hamiltonian [7–10]:

$$H_r(I, \tilde{\varphi}) = -(\omega - 1)I + \epsilon J_1(\rho) \cos(\tilde{\varphi}),$$
$$I = \frac{\tilde{p}^2 + \dot{\tilde{p}}^2}{2}, \qquad \rho = \sqrt{2I},$$
$$\tilde{\varphi} = \varphi - \omega \tilde{t} + \pi, \qquad \varphi = \arctan\left(\frac{\tilde{p}}{\dot{\tilde{p}}}\right),$$
$$\tilde{p} = \rho \sin(\varphi), \qquad \dot{\tilde{p}} = \rho \cos(\varphi), \qquad (12)$$

where $J_1(x)$ is a Bessel function of the first order [29].

If $|\omega - 1|$ is sufficiently small, the Hamiltonian (12) possesses saddles generating separatrices [Figs. 2(a) and 2(b)]. When $\omega = 1$, the separatrices merge into a single infinite grid [Fig. 2(a)]. For the original system (5), the neglected fast-oscillating terms dress this grid with a chaotic layer, thus, forming a stochastic web (SW). Formally, chaotic diffusion along the vertical filaments of the SW might transport the system to arbitrarily high values of $I$, so that $|\vec{p}|$ might become arbitrarily large. In all former works, e.g., Refs. [5,6,11–27], it was this chaotic diffusion that was believed to be the origin of the resonant drift. This cannot be the case, however, because (i) at $\epsilon/4 \ll 1$, the time scale at which chaos manifests [7–10] is much larger than that for the formation of the resonant peak (being $\sim \omega_c^{-1} \min\{\tilde{t}_s, \tilde{t}_{\rm SM}\}$), and (ii) at $\epsilon/4 \gtrsim 1$, when chaos is pronounced, $\vec{p}$ varies chaotically at relevant time scales indeed, but this leads to a chaotic variation of the value and sign of $v_x$ (2) in the integrand of the integral in Eq. (3), which decreases the integral rather than increasing it; therefore, chaos suppresses the drift.

We uncover the true origin of the resonant peak in the general case by an analysis of the regular dynamics along the trajectory of the resonant Hamiltonian (12) starting from $(I = +0, \tilde{\varphi} = \pi/2)$ [28]. In the equations of motion for the system (12), we transform from $I$ to $\rho$ and scale the time and frequency shift by the slow "time" $\tilde{t}_{\rm SM}$ and its reciprocal, respectively,

$$\frac{d\rho}{d\tau} = \frac{J_1(\rho)}{\rho} \sin(\tilde{\varphi}), \qquad \frac{d\tilde{\varphi}}{d\tau} = -\delta + \frac{dJ_1(\rho)}{d\rho} \cos(\tilde{\varphi}),$$
$$\tau \equiv \frac{\tilde{t}}{\tilde{t}_{\rm SM}} \equiv \epsilon \tilde{t}, \qquad \delta \equiv \frac{\omega - 1}{\tilde{t}_{\rm SM}^{-1}} \equiv \frac{\omega - 1}{\epsilon}. \qquad (13)$$

For $|\omega - 1| \ll 1$, the slow dymanics of $\tilde{p}$ is fully described by solution of Eq. (13) with appropriate initial conditions [28]

$$\rho(\tau = 0) = +0, \qquad \tilde{\varphi}(\tau = 0) = \pi/2. \qquad (14)$$

The drift velocity is [28]

$$\tilde{v}_d = \tilde{v}_{\rm ET}(\omega/\tilde{\nu}) + \tilde{v}_d^{\rm (res)}(\delta, \alpha),$$
$$\tilde{v}_d^{\rm (res)} = \frac{\int_0^{\tau_p} d\tau \exp\left(-\frac{\tau}{4\alpha}\right) J_1[\rho(\tau)] \sin[\tilde{\varphi}(\tau)]}{4\alpha[1 - \exp\left(-\frac{\tau_p}{4\alpha}\right)]}, \qquad (15)$$

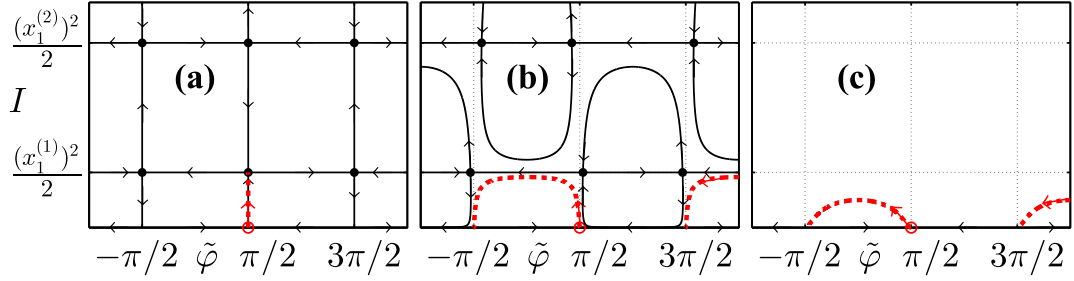

FIG. 2 (color online). Phase plane of the resonant Hamiltonian (12) for three characteristic values of $\delta \equiv (\omega - 1)/\epsilon$: (a) $\delta = 0$, (b) $0 < \delta < \delta_{cr}^{(2)}$, (c) $\delta > \delta_{cr}^{(1)}$, where $\delta_{cr}^{(n)} = \{(1/x)|[dJ_1(x)/dx]|\}|_{x=x_1^{(n)}}$. The $x_1^{(n)}$ are $n$th zeros of the Bessel function $J_1(x)$. Red circles mark the points $(+0, \pi/2)$; the red dashed lines show outgoing trajectories [in (b),(c), the same applies to equivalent trajectories from $(+0, 5\pi/2)$]. Dots mark saddles; separatrices are shown by solid lines. Arrows indicate directions of motion.

where $\tau_p$ is the period of the trajectory (13) and (14). Figure 3 demonstrates the effectiveness of Eq. (15). Figure 3(a) relates to the aforementioned case (11) with $\theta = 40°$: the agreement between the theory and simulations in the range of the resonant peak is excellent. Figure 3(b) shows the evolution of $\tilde{v}_d(\omega)$ in the vicinity of $\omega = 1$ as $\alpha$ grows while $\tilde{\nu} = 0.02$. In addition to perfect agreement for $\epsilon \equiv 4\alpha\tilde{\nu} \lesssim 0.4$ and reasonable agreement for higher $\epsilon$ up to 0.8, it illustrates the key features discussed below.

A striking feature of Fig. 3(b) is the nonmonotonic dependence of the peak maximum on $\alpha$. It is a consequence of an interplay between the time scales $\tilde{t}_s \equiv \tilde{\nu}^{-1}$ and $\tilde{t}_{SM} \equiv \epsilon^{-1}$. To further clarify the role of the latter, consider the exact resonance: $\delta = 0$. The initial $\tilde{\varphi} = \pi/2$ is then preserved along the trajectory (13) and (14), so that $\rho$ obeys the closed dynamical equation $d\rho/d\tau = J_1(\rho)/\rho$. At $\tau \equiv \tilde{t}/\tilde{t}_{SM} \ll 1$, the rhs is equal to $1/2$ [29], so that $\rho$ reaches values $\sim 1$ for $\tau \sim 1$. The growth of $\rho$ then slows down, and, for $\tau \sim 1$, $\rho$ reaches the vicinity of $x_1^{(1)} \approx 4$ corresponding to the first saddle of the SW [Fig. 2(a)], where the growth saturates. Moreover, the stay in the vicinity of the saddle results in that resonant drift ceases. If $\tilde{t}_s \ll \tilde{t}_{SM}$, the saturation of the $\rho$ growth is irrelevant. So, increase of $\epsilon$ results in faster acceleration of the transverse momenta during the whole period before scattering, and, by the time of the scattering, their magnitude has reached higher values; the same applies to the modulation amplitude and, thus, $\tilde{v}_d^{(res)}$ too. In the opposite limit $\tilde{t}_{SM} \ll \tilde{t}_s$, the drift stops at $\tilde{t} \sim \tilde{t}_{SM}$—long before the scattering. In this regime, the probability $P_{RD}$ for electron to undergo the resonant drift is $\sim \tilde{t}_{SM}/\tilde{t}_s$. Since $\tilde{v}_d^{(res)}$ is proportional to $P_{RD}$, it decreases together with $\tilde{t}_{SM} \equiv \epsilon^{-1}$. The optimal regime is $\tilde{t}_s \sim \tilde{t}_{SM}$, i.e., $\alpha \sim 1$.

Figure 4(a) shows the universal function $A(\alpha)$ representing the resonant peak maximum $\tilde{v}_d^{(res)}(\delta = 0, \alpha)$ [the analytic formula is given in Eq. (S.6) of [28]]. It attains the maximum $A_{max} \approx 0.38$ at $\alpha = \alpha_{max} \approx 1.16$ while its small-$\alpha$ and large-$\alpha$ asymptotes are $\alpha$ and $(x_1^{(1)})^2/(8\alpha) \approx 1.84/\alpha$, respectively. Figure 4(b) compares $\tilde{v}_d(\omega = 1)$ and $\tilde{v}_{ET}(1/\tilde{\nu}) + A[\epsilon/(4\tilde{\nu})]$ as functions of $\epsilon$ for a given $\tilde{\nu} = 0.02$. The agreement is excellent up to $\epsilon \approx 0.3$ and good up to $\epsilon \approx 0.7$.

Figure 3(b) demonstrates also that, as $\alpha$ increases, the width of the peak grows monotonically while its shape evolves from being domelike to being spikelike. Analytic results are presented in Ref. [28].

Finally, Fig. 3(b) demonstrates that chaos comes into play only at $\epsilon \sim 1$, leading to fluctuations in $\tilde{v}_d(\omega)$ (see the curve for $\alpha = 10$). As $\epsilon$ increases further, fluctuations intensify while the peak disappears [see the curve for $\alpha = 15$ and the range $\epsilon \gtrsim 1.2$ in Fig. 4(b)]. See Ref. [28] for

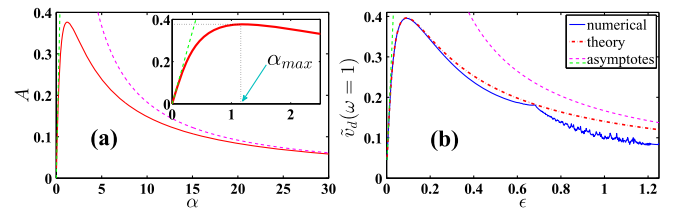

FIG. 4 (color online). (a) Universal asymptotic dependence (S.6) of the amplitude of the resonant peak on $\alpha \equiv \epsilon/(4\tilde{\nu})$ (solid line) and its asymptotes for small and large $\alpha$ (dashed lines). The inset shows the enlarged scale for $\alpha < 2.5$. (b) Comparison between (i) numerically calculated $\tilde{v}_d(\omega = 1)$ for $\tilde{\nu} = 0.02$ as a function of $\epsilon$ (blue thin solid lines) and (ii) the theory—the general theory (red dash-dotted line) i.e. Eq. (15) for $\omega = 1$ while $v_d^{(res)}(\omega = 1)$ reduces to the expression given in Eq. (S.6) of [28], and the small-$\alpha$ or large-$\alpha$ asymptotes (dashed lines) i.e. (15) for $\omega = 1$ while $v_d^{(res)}(\omega = 1)$ is approximated as $\alpha$ or $(x_1^{(1)})^2/(8\alpha)$ respectively.

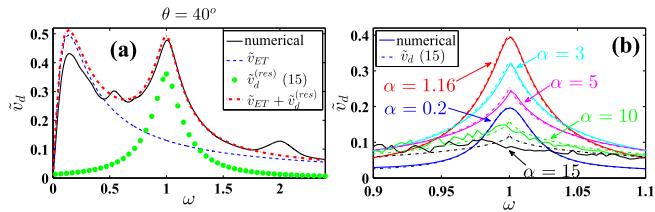

FIG. 3 (color online). Scaled drift velocity vs the ratio between the Bloch and cyclotron frequencies: the general theory (15) and the numerical simulations for (a) the case Eq. (11) with $\theta = 40°$, (b) $\tilde{\nu} = 0.02$ as $\alpha \equiv \epsilon/(4\tilde{\nu})$ increases.

details. Thus, chaos may be relevant only at $\epsilon \gtrsim 1$, playing a destructive role for the resonant drift, contrary to the established belief [5,6,11–27] about its constructive role. The latter belief suggested that the best performance of the resonant drift occurs when chaos is strong. However, neither simulations nor experiments [6,17,18,21] confirm this: as $\theta \to 90°$, when chaos intensifies to its maximum extent, the drift vanishes. Our work shows that the ways needed to control the resonant drift are different. If the model (1) is valid and $\tilde{\nu} \ll 1$ and $\epsilon/4 \ll 1$, it is controlled by a single parameter $\alpha$. The best performance corresponds to $\alpha = \alpha_{max} \approx 1.16$. The drift is negligible if any of the following conditions hold: $\alpha \lesssim 0.1$, $\alpha \gtrsim 20$, $\tilde{\nu} \gtrsim 1$. As $\epsilon$ grows above 1, the resonant drift at $\omega \approx 1$ gradually decays (at multiples, it first rises and then decays too). See Ref. [28] for illustrations.

In conclusion, we have shown that the enhancement of the electron drift occurring if the Bloch and cyclotron frequencies are close, originates in a regular dynamics, contrary to the widespread belief that its origin is in chaotic diffusion. The enhancement is explained as follows. The electron motion along the SL and the tilted magnetic field produce a Lorentz force oscillating at the Bloch frequency. It excites cyclotron rotation which modulates the angle of the Bloch oscillation of the instantaneous velocity. Beyond resonance, the velocity change caused by the modulation oscillates during the relevant time scale, and so the drift averages to zero. In contrast, the change in the resonant case keeps its sign, thus, being accumulated.

Discussions with Vladimir Kravtsov, Oleg Yevtushenko, Boris Glavin, and Oleg Raichev are much appreciated. S. M. S. acknowledges the CMSP section of the Abdus Salam International Centre for Theoretical Physics for the support of his visit there when a part of the Letter was prepared. The research was supported in part by the Engineering and Physical Sciences Research Council UK (Grant No. EP/G070660/1).

# SUPPLEMENTARY MATERIAL
## for
## "Regular Rather than Chaotic Origin of the Resonant Transport in Superlattices"

S. M. Soskin, I. A. Khovanov, P. V. E. McClintock

March 26, 2015

### SUMMARY

We provide, for a subset of interested specialists, more-detailed derivations of some of the main equations of the Letter together with additional discussions and illustrative figures.

# Contents

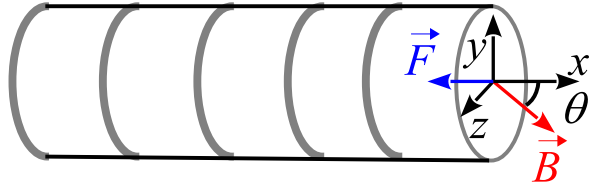

Figure S.1: Schematic representation of the superlattice, electric field $\vec{F}$, magnetic field $\vec{B}$, and coordinate axes.

# 1 Physical origin of the dynamics

In this section, we discuss and explain the physical nature of the dynamics (5) and its effect on the instantaneous velocity. To this end, we show a schematic view of the SL (Fig. S.1) and present Eq. (1) in a more explicit form – as a system of dynamical equations for the components of $\vec{p}$ –

$$\dot{p}_x = eF - \omega_\perp p_y, \qquad \text{(S.1)}$$
$$\dot{p}_y = -\omega_\parallel p_z + \omega_\perp \tilde{v}_x,$$
$$\dot{p}_z = \omega_\parallel p_y,$$

$$\tilde{v}_x \equiv v_x m^* = \frac{d \Delta m^*}{2\hbar} \sin\left(\frac{p_x d}{\hbar}\right), \qquad \omega_\perp \equiv \frac{eB}{m^*} \sin(\theta), \qquad \omega_\parallel \equiv \frac{eB}{m^*} \cos(\theta).$$

If there was no motion along the SL, i.e. if $\tilde{v}_x = 0$, the 2nd and 3rd equations would merely correspond to cyclotron rotation in the transverse plane (i.e. $y - z$) under the influence of the magnetic field[*]. On the other hand, such an autonomous rotation is possible only if the momentum in the transverse plane is non-zero, which is not in fact the case at the initial instant: $p_y(0) = p_z(0) = 0$ (see (7) and Sec. 2 below). However, as time goes by, electron motion along the SL does occur which, given the $z$ component of the magnetic field, results in a Lorentz force causing the electron to move in the $y$ direction, in turn triggering cyclotron rotation in the transverse plane. At the same time, the onset of motion in the $y$ direction plays another important role: together with the $z$ component of the magnetic field, it generates a Lorentz force in the $x$ direction which, in turn, changes $p_x$ (see the 2nd term on the rhs of the 1st equation of (S.1)). The latter causes a change of the instantaneous velocity $v_x(t)$ through the change of its angle (i.e. the argument of the sine in the definition of $\tilde{v}_x$ in (S.1)). In a sense, this complicated Bloch "oscillation"[†] dynamics is a consequence of the specific two-stage feedback.

To complete this physical picture of motion in the system, we note that those variations of $p_x$ and $p_z$ in time which relate to the magnetic field are mutually correlated because they are caused respectively by the $x$ and $z$ components of one and the same Lorentz force (generated by the magnetic field and the electron motion in the $y$ direction): these components therefore vary in time coherently (cf. the 2nd term on the rhs of the 1st equation with the rhs of the 3rd equation in (S.1)). That is why $p_x$ possesses, not only the conventional component proportional to $t$ (which is what leads to the Bloch oscillations), but also a component proportional to $p_z$. Substituting the expression for $p_x$ in terms of $t$ and $p_z$ into the argument of the sine in $\tilde{v}_x$, and using the second of the dynamical equations (S.1), we see that the dynamics of $p_z$ and $p_y$ reduces to cyclotron rotation perturbed by a wave-like term at the Bloch frequency or, equivalently, to the dynamics (5).

---

[*]The frequency of such a rotation is entirely determined by the properties of electron (charge and effective mass) and by a value of the $x$ component of $\vec{B}$.

[†]Obviously, the term "oscillation" is not fully adequate in this case since the angle of the "oscillation" is affected by one of the dynamical variables. But we still use this term for the sake of brevity.

We conclude that, in the presence of a magnetic field in the $x - z$ plane, electron motion along the SL axis (the Bloch oscillation) excites cyclotron rotation in the transverse ($y - z$) plane which, in turn, provides for a feedback on the motion along the SL through modulation of the angle of the velocity $v_x(t)$. In mathematical terms, it is expressed in Eqs. (5) and (2) together with the relationship between $\tilde{p}_x$ and $\tilde{p}_z \equiv \tilde{p}$ which, for the relevant initial conditions (7), is $\tilde{p}_x(\tilde{t}) = \omega \tilde{t} - \tilde{p}(\tilde{t})$.

## 2  Initial conditions

We now consider the relevance of the initial conditions (7), which are equivalent to zero[1,2,3] initial values of the quasi-momenta:

$$p_x(0) = p_y(0) = p_z(0) = 0, \tag{S.2}$$

and then derive on the base of these conditions the initial conditions (14) for the slow variables $(\rho, \tilde{\varphi})$.

The crystal momenta of the electrons contributing to the electric current are typically related to the Fermi energy, and so are non-zero even at the absolute zero of temperature[4]. So, why do we use (S.2) then? The answer is as follows.

Let us first clarify the physical meanings of $p_x$, $p_y$ and $p_z$. The quantity $p_x$ is the SL analogue of crystal momentum in a bulk crystal. The meanings of $p_y$ and $p_z$ are rather different. Not only do they relate to the bulk semiconductor rather than to the SL, but they also represent a *deviation* of the crystal momenta from their values at the bottom of the conduction band – which is the essence of the effective mass approximation[4].

Now consider the relationship between temperature and the initial momenta. Before moving further we note that, though we use the term "zero temperature", we actually assume a temperature that is very low but non-zero, consistent with real experiments where the temperature is typically the boiling point of liquid-He 4.2 K. Strictly at $T = 0$ the semiconductor would of course be an insulator, and any transport would be impossible.

Consider first $p_x$. Typically, the Fermi energy in the emitter contact falls below the bottom of the relevant (lowest) miniband of the SL[5,6,2]. But some electrons will be injected from the emitter into the SL nevertheless, since temperature is non-zero. Moreover, the relevant SLs are designed in such a way that the bottom of the miniband is lowered almost to the Fermi energy in the emitter[7,6] so that the injection of electrons is greatly eased in comparison with the conventional case[5]. On the other hand, before being affected by the external fields, the electrons in the SL fill only a small range of energies $\sim k_B T$ above the bottom of the miniband where $k_B$ is Boltzmann's constant (note that $k_B T \ll \Delta$). It means that the scaled initial momentum $\tilde{p}_{x0} \equiv p_x(0)d/\hbar$ (5) is distributed mostly in the range $\sim \sqrt{k_B T/\Delta}$ around zero. As $T \to 0$, this quantity goes to zero[‡].

---

[‡]E.g. for the SLs used in[6,8,9] and for $T = 4.2$ K, the corresponding width of the $\phi_0$ distribution is $\sim 10^{-3}$ and thus is extremely small compared to the relevant scale $2\pi$

Similarly, for transverse motion the number of electrons in the conduction band is small provided $T$ is sufficiently small. Therefore the initial "energies" $p_y^2(0)/(2m^*)$ and $p_z^2(0)/(2m^*)$ are distributed within the range $\sim k_B T$. Accordingly, the absolute values of the scaled initial momenta $\tilde{p}_{y0} \equiv p_y(0)d/\hbar$ and $\tilde{p}_{z0} \equiv p_z(0)d\tan(\theta)/\hbar$ are $\sim d\sqrt{m^* k_B T}/\hbar$ and $\sim d\tan(\theta)\sqrt{m^* k_B T}/\hbar$ respectively. In the asymptotic limit $T \to 0$, the ratio of these quantities to the characteristic dynamical scale in the problem, being $\sim \min(\alpha, \alpha_{\max})$, vanishes[§]. That is why the initial momenta $p_y(0)$ and $p_z(0)$ may be set to $0$ in the asymptotic limit $T \to 0$.

Finally, we note that, for SL parameter values relevant to the experiments, the above qualitative estimates conform with the numerical calculations in[3] for the temperature dependence of $\tilde{v}_d(\omega)$ and with the generalization of our analytic results to the case of non-zero temperatures (the results will be presented elsewhere): in the range $T \lesssim 4$ K, the function $\tilde{v}_d(\omega)$ is almost indistinguishable from that for $T = 0$.

We now turn to the dynamics (13) in slow polar coordinates $\rho - \tilde{\varphi}$ (12). Due to the zero initial conditions (7) for the momenta, the relevant initial $\rho$ for the system (13) is $\rho = 0$. In order to avoid a singularity on the rhs of the equation for $d\tilde{\varphi}/dt$, it is necessary to take an infinitesimal positive value instead, which we denote as $+0$. The initial angle $\tilde{\varphi}$ is formally indefinite. However, if $\rho = +0$, then, for any $\tilde{\varphi}$ from the ranges $]-\pi/2, \pi/2[$ and $]\pi/2, 3\pi/2[$, the absolute value of the derivative $d\tilde{\varphi}/dt$ diverges with a sign that is positive or negative respectively. Thus, the system (13) with an initial $\rho$ equal to $+0$ and an initial $\tilde{\varphi}$ lying beyond an infinitesimal vicinity of the value $-\pi/2$ is immediately transferred to an infinitesimal vicinity of the point $(\rho = +0, \tilde{\varphi} = \pi/2)$, from which motion starts with finite derivatives of both variables. That is why the relevant initial conditions are those given in Eq. (14).

## 3 Amplitude of the peak

The maximum of the drift velocity $\tilde{v}_d$ as function of frequency $\omega$ occurs at the exact resonance, i.e. when $\delta = 0$ (see Sec. 4 below). In this case, the Hamiltonian system (12) possesses a grid-like separatrix and the system moves along its first vertical segment at $\tilde{\varphi} = \pi/2$ (Fig. 2(a)). In terms of the polar coordinates $\rho - \tilde{\varphi}$, it corresponds to $\tilde{\varphi} = \pi/2$ being preserved while $\rho$ moving accordingly from $+0$ towards the first zero of the Bessel function of the 1st order $x_1^{(1)} \approx 3.83$[10], that corresponds to the lowest-action saddle point of the separatrix of the Hamiltonian system (12). Indeed, if $\delta = 0$ while $\tilde{\varphi}(0) = \pi/2$, then the latter is preserved at the trajectory (13)-(14) as follows from the second equation in (13). Due to this, the dynamics of $\rho(\tau)$ reduces to the following closed dynamic equation:

$$\frac{d\rho}{d\tau} = \frac{J_1(\rho)}{\rho}, \qquad (S.3)$$

which is immediately integrated in quadratures. Thus, taking account of $\rho(0) = +0$,

$$\tau = \int_{+0}^{\rho} dx \frac{x}{J_1(x)}. \qquad (S.4)$$

Expressing $\tilde{p}$ in terms of the polar coordinates (12), i.e. $\tilde{p} = \rho \sin(\varphi) \equiv \rho \sin(\tilde{\varphi} - \pi + \omega \tilde{t})$, and allowing for $\tilde{\varphi} = \pi/2$ and $\omega = 1$, we obtain:

$$\tilde{p} = -\rho \cos(\tilde{t}). \qquad (S.5)$$

---

[§]For example, for the experiments[6,8,9] where typical values of $\alpha$ and $\tan(\theta)$ where $\sim 1$, the ratios were $\sim 0.1$ in case of $T = 4.2$ K

Substituting this expression into the equations for the drift velocity (6) with $\omega = 1$, using the trivial formula $\sin(\tilde{t} - \tilde{p}) = \sin(\tilde{t})\cos(\tilde{p}) - \cos(\tilde{t})\sin(\tilde{p})$, allowing for Fourier expansions of the functions $\cos(\rho\cos(\tilde{t}))$ and $\sin(\rho\cos(\tilde{t}))$[10] (the Fourier coefficients are expressed in terms of Bessel functions), using the solution (S.4) and keeping only terms of lowest order in $\epsilon$ and $\tilde{\nu}$, we can derive ¶:

$$\tilde{v}_d(\omega = 1) = \tilde{v}_{ET}(1/\tilde{\nu}) + \tilde{v}_d^{(res)}(\delta = 0, \alpha), \quad (S.6)$$
$$\tilde{v}_d^{(res)}(\delta = 0, \alpha) \equiv$$
$$\equiv A(\alpha) = \frac{1}{4\alpha} \int_{+0}^{x_1^{(1)}} d\rho \exp\left(-\frac{\gamma(\rho)}{4\alpha}\right) \rho,$$
$$\gamma(\rho) \equiv \int_{+0}^{\rho} dx \frac{x}{J_1(x)}, \qquad x_1^{(1)} \approx 3.83,$$

where $\alpha$ is defined in (8).

The small-$\alpha$ and large-$\alpha$ asymptotes of $A(\alpha)$ are:

$$\tilde{v}_d^{(res)}(\delta = 0, \alpha \to 0) \equiv A_0(\alpha) = \alpha, \quad (S.7)$$
$$\tilde{v}_d^{(res)}(\delta = 0, \alpha \to \infty) \equiv A_\infty(\alpha) = \frac{\left(x_1^{(1)}\right)^2}{8\alpha} \approx \frac{1.84}{\alpha}.$$

A graphical presentation of the function $A(\alpha)$ with the asymptotes is provided in the main paper: see Fig. 4 and the associated discussion.

## 4 Full peak

Consider the case of inexact resonance, i.e. $\delta \neq 0$ (but, if $\delta \to 0$, then $\tilde{v}_d^{(res)}$ obviously reduces to that for $\delta = 0$). Similarly to the case $\delta = 0$, one can show that $\tilde{v}_d$ may be presented as a superposition of the Esaki-Tsu peak and the resonant contribution (which, as function of $\delta$, takes the form of a peak) while, for a given value of $\delta$, the latter depends only on $\alpha$. The resonant contribution is significant only in the vicinity of the resonance (i.e. where $|\omega - 1| \equiv |4\delta\alpha\tilde{\nu}| \ll 1$), where it is given by the following semi-explicit formula:

$$\tilde{v}_d^{(res)} = \frac{1}{4\alpha} \int_0^\infty d\tau \exp\left(-\frac{\tau}{4\alpha}\right) J_1(\rho(\tau)) \sin(\tilde{\varphi}(\tau)), \quad (S.8)$$

where $\rho(\tau)$ and $\tilde{\varphi}(\tau)$ are solutions of the system of dynamical equations (13) with the initial conditions (14).

One may simplify the formula (S.8), thereby easing both the calculation of $\tilde{v}_d^{(res)}$ and its analysis. To do so, we present a typical trajectory (13)-(14) for $\delta > 0$ ‖ in Fig. S.2 and the corresponding functions $J_1(\rho(\tau))$, $\sin(\tilde{\varphi}(\tau))$ and their product in Fig. S.3. All functions shown in Fig. S.3 are periodic. Their period $\tau_p$ is equal to the time of motion along the trajectory, that in turn is equal to the time of motion from the point $(\rho = +0, \tilde{\varphi} = \pi/2)$ to the point $(\rho =$ $+0$, $\tilde{\varphi} = -\pi/2$) because the time spent by the system at the segment of the trajectory lying at an infinitesimal height above the $\tilde{\varphi}$-axis in between the values of $\tilde{\varphi}$ equal to $-\pi/2$ and $-3\pi/2$ is infinitesimal. Let us divide the range of integration in the integral (S.8) for intervals $[0, \tau_p]$, $[\tau_p, 2\tau_p]$, $[2\tau_p, 3\tau_p]$,... Presenting the integral (S.8) as a sum of integrals over these intervals, allowing for the fact that the quantity $J_1(\rho(\tau)) \sin(\tilde{\varphi}(\tau))$ entering the integrand is periodic, using the multiplicative property of the exponential function (i.e. $\exp(a + b) = \exp(a)\exp(b)$) and the formula for the sum $\sum_{n=0}^{\infty} \exp\left(-\frac{n\tau_p}{4\alpha}\right) = 1/\left(1 - \exp\left(-\frac{\tau_p}{4\alpha}\right)\right)$, we ultimately obtain Eq. (15), reproduced below for readers' convenience:

$$\tilde{v}_d^{(res)} \equiv \tilde{v}_d^{(res)}(\delta, \alpha) = \frac{\int_0^{\tau_p} d\tau \exp\left(-\frac{\tau}{4\alpha}\right) J_1(\rho(\tau)) \sin(\tilde{\varphi}(\tau))}{4\alpha \left(1 - \exp\left(-\frac{\tau_p}{4\alpha}\right)\right)}, \tag{S.9}$$

where $(\rho(\tau), \tilde{\varphi}(\tau))$ is the solution of the system of dynamical equations (13) with the initial conditions (14), while $\tau_p$ is its period. Note that, for a given $\alpha$, this solution (and its period, in particular) depends only on the parameter $\delta$. In the cases of $\delta = 0$ or $\delta \gg 1$, it can be found in quadratures or explicitly, respectively. In the general case of an arbitrary $\delta$, it can easily be found numerically.

---

¶As for the term $\tilde{v}_{ET}(1/\tilde{\nu}) \approx \tilde{\nu}$ (that may formally be attributed to the contribution from the Esaki-Tsu peak defined in Eq. (4)), it was derived using the following properties, that can be rather easily proved. If there is an arbitrary function $f(x)$ and the scale of $x$ at which it significantly changes greatly exceeds 1, while $f(0) = 1$ and $f(\infty) = 0$, then the integral $I \equiv \int_0^\infty dx \sin(x) f(x)$ is approximately equal to 1. Similarly, if $f(0) = 0$, then $I \approx 0$.

‖The case $\delta < 0$ differs only by that the trajectory symmetrically goes in the opposite direction.

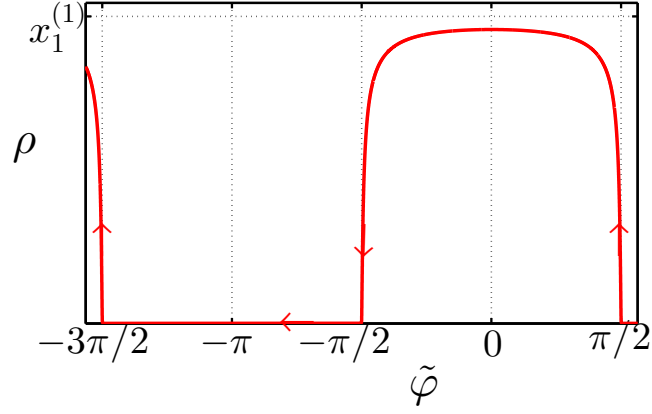

Figure S.2: A typical trajectory (13)-(14) (red solid line). Arrows show the direction of motion.

Eq. (S.9) gives a complete quantitative description of the resonant peak of the drift velocity in the asymptotic limit $\tilde{\nu} \to 0$. It is easy to see that $\tilde{v}_d^{(res)}$ is an even function of $\delta$. As $\delta \to 0$, the period $\tau_p$ diverges, the maximum of the quantity $J_1(\rho(\tau)) \sin(\tilde{\varphi}(\tau))$ approaches its maximum possible value ($\max(J_1) \approx 0.58$[10]), and the width of the peak of this function of $\tau$ also approaches its maximum. Altogether, this means that $\tilde{v}_d^{(res)}$ as function of $\delta$ attains its maximum at $\delta = 0$, which is described in Sec. 3 above [**].

We note also that, in the asymptotic limits $\alpha \to 0$ and $\alpha \to \infty$, the expression (S.9) significantly simplifies:

1. For $\alpha \to 0$, (S.9) can be shown to reduce to $\tilde{v}_{d,a}^{(res)}(\delta, \alpha)$ (10) in the asymptotic limit $\tilde{\nu} \to 0$:

$$\tilde{v}_d^{(res)}(\delta, \alpha \to 0) = A_0(\alpha) L(4\delta\alpha), \tag{S.10}$$

$$A_0(\alpha) \equiv \alpha, \qquad L(x) \equiv \frac{1}{1 + x^2}, \qquad 4\delta\alpha \equiv \frac{\omega - 1}{\tilde{\nu}}.$$

Here, $L(x)$ is a conventional Lorentzian.

---
[**]It is rather easy to show that, for $\delta \to 0$, the formula (S.9) reduces to (S.6).

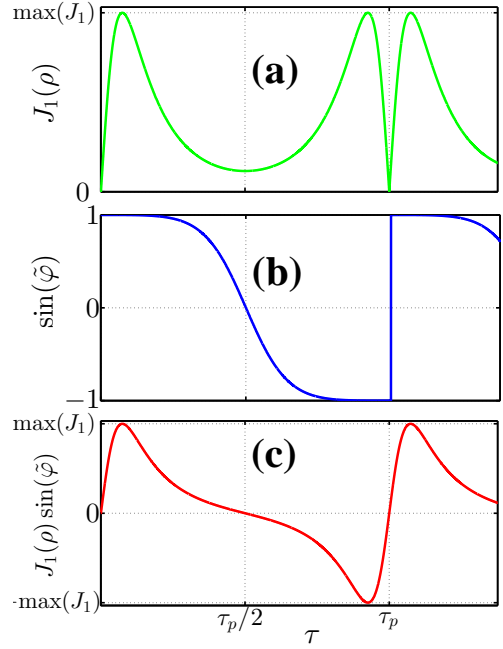

Figure S.3: Functions corresponding to the trajectory shown in Fig. S.2: (a) $J_1(\rho(\tau))$, (b) $\sin(\tilde{\varphi}(\tau))$, and (c) $J_1(\rho(\tau))\sin(\tilde{\varphi}(\tau))$ (note that $\max(J_1) \approx 0.58^{10}$).

2. For $\alpha \to \infty$, (S.9) can be shown to simplify as follows:

$$\tilde{v}_d^{(res)}(\delta, \alpha \to \infty) = A_\infty(\alpha) K(\delta), \tag{S.11}$$

$$A_\infty(\alpha) \equiv \frac{\left(x_1^{(1)}\right)^2}{8\alpha}, \qquad K(\delta) \equiv \frac{\int_0^{\tau_p} d\tau (\tau_p/2 - \tau) J_1(\rho(\tau)) \sin(\tilde{\varphi}(\tau))}{\left(x_1^{(1)}\right)^2 \tau_p}.$$

Here, $K(x)$ is a universal function which, to the best of our knowledge, has not been studied earlier in any context and which we will refer to below as the $K$-form. Like the Lorentzian, it is an even function, being equal to 1 at $x = 0$, but its other features are very different: it has a very sharp spike-like maximum (cf. the smooth dome-like maximum of the Lorentzian) while its far wings decay as slowly as those of a Lorentzian i.e. $\propto x^{-2}$ (Fig. S.4).

Let us return to the general formula (S.9). Figs. 3(b) and 4(a) demonstrate that, as $\alpha$ grows, the amplitude of the peak of $\tilde{v}_{d,a}^{(res)}$ at first increases, attaining a maximum at $\alpha = \alpha_{\min} \approx 1.16$, and then gradually decreases to zero. If $\tilde{\nu}$ is fixed, the half-width at half-maximum, expressed in terms of $\omega$, monotonically grows with $\alpha$ – from $\tilde{\nu}$ at $\alpha \ll 1$ to values comparable with 1 at $\alpha \gtrsim 1/(4\tilde{\nu})$ (Figs. S.5 and S.6). The shape evolves very substantially – from the dome-like Lorentzian at small $\alpha$ to the stretched $K$-form at large $\alpha$. Fig. S.5 illustrates the evolution of the resonant peak for which the magnitude and the width are scaled by the amplitude $A(\alpha)$ and $\tilde{\nu}$ respectively.

To characterize the universal evolution of the peak shape quantitatively, we proceed as follows. For any given value of $\alpha$, the resonant peak is fully described by: (i) an amplitude; (ii) a half-width (HW); and (iii) a doubly-scaled shape. The universal dependence of the amplitude, $A(\alpha)$, has been already found (Eq. (S.6) and Fig. 4(a)). The HW depends on (a) the factor $r$ characterizing the drop in the height of the peak at the chosen HW, as compared to that of the maximum, and (b) the units in which the HW is expressed. For (a), we use two characteristic examples for the ratio $r$. One is $1/2$ (which is the commonest choice), and the other is $1/3$. For (b), the most reasonable candidates are the following: $\omega - 1$, $(\omega - 1)/\epsilon \equiv \delta$ and $(\omega - 1)/\tilde{\nu} \equiv 4\delta\alpha$. The quantity $\omega - 1$ does not suit us because our purpose is to present the evolution in a universal form. Of the two remaining candidates, we choose $(\omega - 1)/\tilde{\nu}$ (like for $y$ in $S(y)$ in Fig. S.5), for the following reason. The evolution is presented by us for increasing $\alpha$, so that the evolution of the HW in the units $(\omega - 1)/\tilde{\nu}$ starts from a finite value while that in the units $(\omega - 1)/\epsilon$ starts from $\infty$. It is therefore natural to choose the former option.

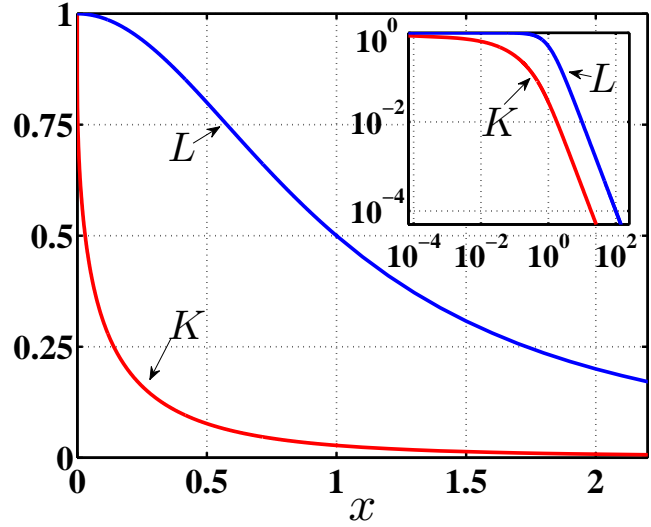

Figure S.4: Comparison of the universal function $K(x)$ (S.11) (red line) with the Lorentzian $L(x)$ (S.10) (blue line). The inset shows the same comparison plotted with logarithmic scales.

So, let us present the resonant contribution $\tilde{v}_d^{(res)}$ (S.9) in the following form:

$$\tilde{v}_d^{(res)} = A(\alpha)\tilde{S}_r(z, \alpha), \tag{S.12}$$

where the doubly-scaled shape $\tilde{S}_r$ is defined as

$$\tilde{S}_r(z, \alpha) = S(y = z\Delta_r(\alpha), \alpha), \tag{S.13}$$

$$S(y, \alpha) \equiv \frac{\tilde{v}_d^{(res)}\left(\delta = \frac{y}{4\alpha}, \alpha\right)}{A(\alpha)},$$

where the HW $\Delta_r$ is defined by the equation

$$S(y = \Delta_r(\alpha), \alpha) = r, \tag{S.14}$$

while $r$ is a value corresponding to the definition of the HW.

The convenience of the presentation (S.12)-(S.14) is that the doubly-scaled shape $\tilde{S}_r(z, \alpha)$ is invariant towards a change of the units in which the half-width is measured.

The universal dependences $\Delta_{1/2}(\alpha)$ and $\Delta_{1/3}(\alpha)$ are shown in Fig. S.6. The HW $\Delta_r$ starts from the value $\sqrt{r^{-1} - 1}$ at $\alpha = 0$ and monotonically increases with $\alpha$, approaching the asymptote $4\Delta_r^{(K)}\alpha$ at very large values of $\alpha$ where the quantity $\Delta_r^{(K)}$ corresponds to a HW of the $K$-form (S.11), defined by the equation $K(\Delta_r^{(K)}) = r$,

$$\Delta_r(\alpha \to 0) = \sqrt{r^{-1} - 1}, \tag{S.15}$$

$$\Delta_r(\alpha \to \infty) \to 4\Delta_r^{(K)}\alpha, \quad K(\Delta_r^{(K)}) = r.$$

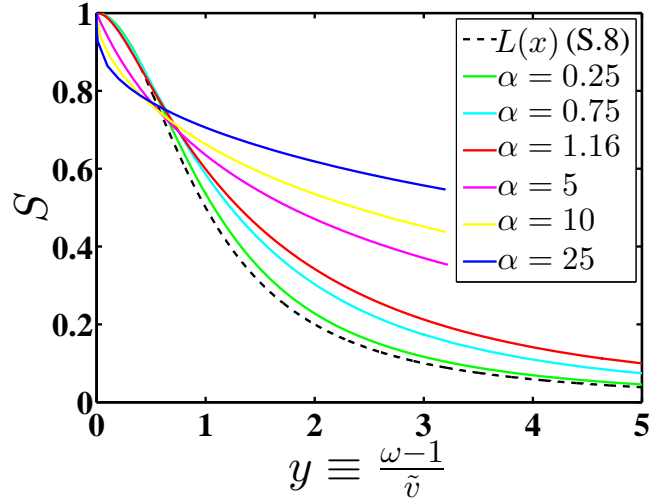

Figure S.5: Evolution of the asymptotic resonant peak scaled by its amplitude, $S(y, \alpha) \equiv \frac{\tilde{v}_d^{(res)}\left(\delta=\frac{y}{4\alpha},\alpha\right)}{A(\alpha)}$, as $\alpha$ grows. The abscissa represents the frequency shift from the resonance scaled by $\tilde{\nu}$: $y \equiv \frac{\omega-1}{\tilde{\nu}}$.

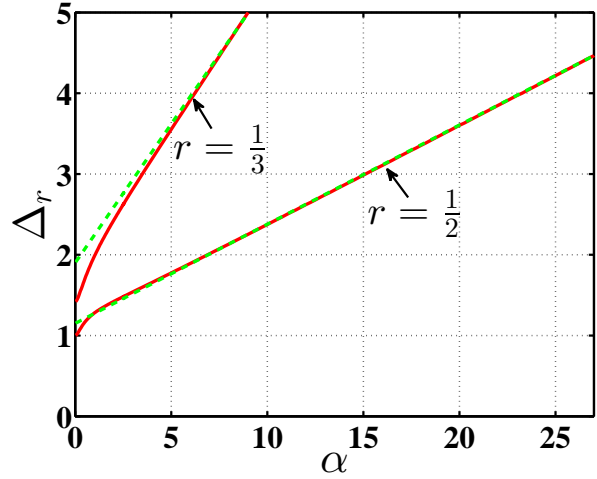

Figure S.6: Half-width $\Delta_r$ (defined by Eq. (S.14)) vs. $\alpha$ for two values of $r$ (namely, $1/2$ and $1/3$), shown by solid lines. The empirical asymptotes (S.16) are shown by dashed lines.

For $r$ equal to $1/2$ and $1/3$, the quantity $\Delta_r^{(K)}$ is equal approximately to $0.031$ and $0.086$ respectively. Thus, a decrease of $r$ by about $30\%$ results in an increase of $\Delta_r^{(K)}$ by factor of more than $200\%$: this reflects the fact that the maximum of the $K$-form is very sharp (Fig. S.4).

It is also worth noting that, for moderate and moderately large values of $\alpha$, the function $\Delta_r(\alpha)$ is well approximated by a straight line parallel to the large-$\alpha$ asymptote but lifted by $\sim \Delta_r(0)$, at least for $r \sim 1/2$ (which is the most important range of $r$):

$$\Delta_{1/2} \approx 1.15 + 4\Delta_{1/2}^{(K)}\alpha, \qquad 1 \lesssim \alpha \lesssim 100, \qquad (S.16)$$
$$\Delta_{1/3} \approx 1.9 + 4\Delta_{1/3}^{(K)}\alpha, \qquad 2 \lesssim \alpha \lesssim 50.$$

As for the invariant doubly-scaled shape $\tilde{S}_r(z, \alpha)$, it evolves from $L\left(z\sqrt{r^{-1}-1}\right)$ to $K(z\Delta_r^{(K)})$ as $\alpha$ increases from $0$ to $\infty$. Such an evolution for $r$ equal to $1/2$ and $1/3$ is shown in Figs. S.7(a) and S.7(b) respectively. Thus, we may conclude: (i) unlike the amplitude, the width and the invariant doubly-scaled shape of the peak evolve monotonically as $\alpha$ increases from $0$ to $\infty$, and (ii) in terms of $\omega$, the HW increases from $\sqrt{r^{-1}-1}\,\tilde{\nu}$ at small $\epsilon$ to $\Delta_r^{(K)}\epsilon$ at $\epsilon \gg \tilde{\nu}$, thus being $\sim \Delta_r^{(K)}$ at $\epsilon \sim 1$.

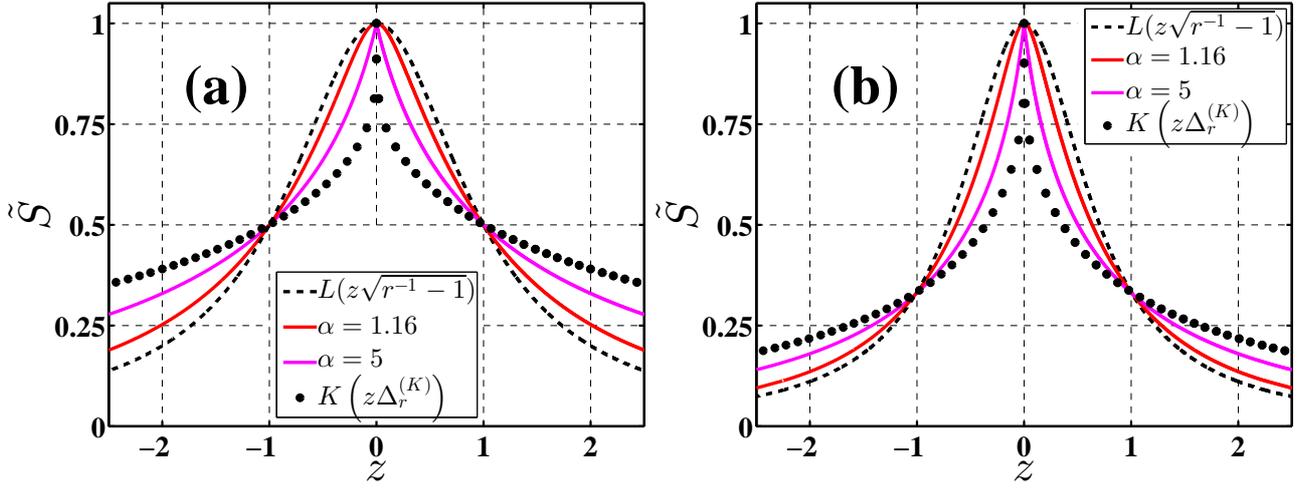

Figure S.7: Evolution of the invariant (doubly-scaled) shape $\tilde{S}_r(x, \alpha)$ (S.13) (solid lines) as $\alpha$ grows, for two values of $r$: (a) $1/2$, and (b) $1/3$. The limiting shapes for $\alpha \to 0$ and $\alpha \to \infty$ are shown by the dashed and dotted lines respectively. The argument represents the ratio of a given deviation from the resonance to its half-width.

## 5 Role of chaos.

It is stated and explained briefly in the main text that, if $\epsilon \ll 1$, the resonant peak is unaffected by chaos even when the latter exists within the system. We now provide some further details.

When $\epsilon \ll 1$, the chaotic layer exists only if $|\omega - 1| \lesssim \epsilon$. Even in the most pronounced case when the resonance is exact (i.e. $\omega = 1$), so that a stochastic web (SW) arises, the chaotic layer forming the SW is extremely narrow[11,12,13,14]: its width is $\propto \exp\left(-\frac{\text{const}\sim 1}{\epsilon}\right)$, so that the time-scale on which chaos manifests (being proportional to the logarithm of the reciprocal of the width) reads in units of $\tau \equiv \epsilon \tilde{t}$ as $\tau_{chaos} = \epsilon^{-1}$. It is explained in the main text that, if $\epsilon \ll 1$, the scale of time in units of $\tilde{t}$ relevant to the formation of the peak is equal to $\min\left(\tilde{\nu}^{-1}, \epsilon^{-1}\right)$. In terms of $\tau$, it is equivalent to $\min(\alpha, 1)$, which is being much smaller than $\tau_{chaos}$. So, even if the relevant trajectory of the system (5) lay within the chaotic layer, its dynamics would be almost regular on the time-scales relevant to the formation of the resonance peak. But moreover, as is known from computer simulations of the equation of motion (5) with zero initial conditions $\left(\tilde{p}=0, \dot{\tilde{p}}=0\right)$, this trajectory lies *beyond the chaotic layer* if $\epsilon$ is sufficiently small[15]. This is explicitly demonstrated in Figs. S.8 (a,b,c).

As $\epsilon$ grows, this regular trajectory gradually deviates more and more from the separatrix of the resonance Hamiltonian (12). The deviation remains relatively small even when $\epsilon$ is only moderately small (Fig. S.8(c)), so that the drift velocity $\tilde{v}_d$ is still well described by Eq. (15) (calculated using the resonant approximation): see Figs. 3 and 4(b).

The trajectory is absorbed by the chaotic layer only if $\epsilon \geq \epsilon_{cr} \approx 0.45$: cf. the plates (c) and (d) in Fig. S.8. If $\epsilon$ only slightly exceeds $\epsilon_{cr}$ (as in Fig. S.8(d)), the trajectory behaves almost regularly for a long time due to sticking to the boundary between the chaotic and non-chaotic areas in the Poincaré section[14]. Thus, $\tilde{v}_d$ is still reasonably described by (15): cf. Fig. 4(b).

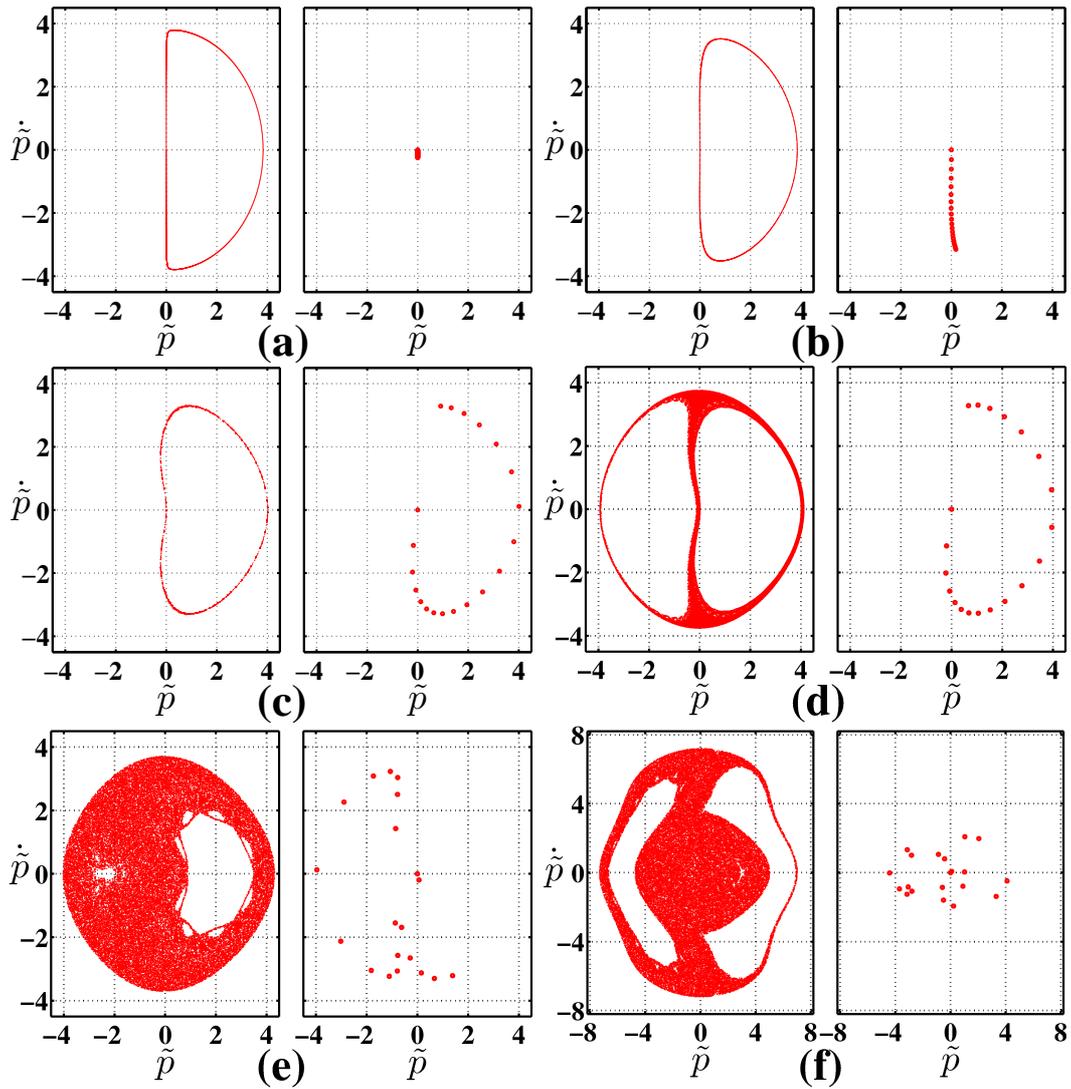

Figure S.8: Poincaré section for the trajectory (5) with $\omega = 1$ and the initial conditions (7) for $50000$ periods of the wave i.e. for the time $\tilde{t} = 100000\pi$ (left panel) and $20$ periods (right panel), as $\epsilon$ grows as: (a) $0.0042$, (b) $0.102$, (c) $0.442$, (d) $0.462$, (e) $0.802$, (f) $2.162$ (note the change of scales in (f)). The number of points in the Poincaré sections in the right-hand panels corresponds to the dynamics during the time $\tilde{t} = 2.5\tilde{\nu}^{-1}$ for $\tilde{\nu} = 0.02$.

As the excess of $\epsilon$ over $\epsilon_{cr}$ grows, the area within boundaries of the chaotic layer sharply grows (cf. the left panels of the plates (d) and (e) in Fig. S.8) while the dynamics becomes more and more chaotic at relevant times (see the right panel of the plate (e) in Fig. S.8). As a result, $\tilde{v}_d(\omega)$ fluctuates as $\omega$ varies (see the curve for $\alpha = 10$ in Fig. 3(b) corresponding to $\epsilon = 0.8$) while its magnitude drops with the increasing $\epsilon$ more sharply than predicted by (S.6) obtained within the resonant approximation: see Fig. 4(b).

Finally, as $\epsilon$ significantly increases further, chaos within the relevant area of the Poincaré section becomes practically global (see the left panel of the plate (f) in Fig. S.8) while the dynamics becomes extremely chaotic (see the right panel of the plate (f) in Fig. S.8) so that the concept of the resonant peak is no longer meaningful: $\tilde{v}_d(\omega)$ fluctuates strongly and it is not possible to identify any characteristic peaks within it.

# 6 Control over the drift

A remarkable result emerging from the research is that, despite the large number of physical parameters affecting resonant single-electron drift, its scaled characteristics entirely depend *only* on a single combination of these parameters,

$$\alpha = \frac{\Delta d^2 eB \sin^2(\theta)}{8\hbar^2 \nu \cos(\theta)}, \tag{S.17}$$

provided that two rather weak conditions hold:

$$\tilde{\nu} \equiv \frac{\nu m^*}{eB \cos(\theta)} \ll 1, \qquad \epsilon \equiv \frac{\Delta m^* d^2 \tan^2(\theta)}{2\hbar^2} \lesssim 1, \tag{S.18}$$

and that the model (1) is valid.

As follows from the Letter and Sec. 4 above, it is natural to define the resonant contribution to the drift at a given value of $\omega$ as the excess of $\tilde{v}_d(\omega)$ (numerically calculated by (5)-(7)) over that in the absence of the magnetic field, i.e. over the Esaki-Tsu function $\tilde{v}_{ET}(\omega/\tilde{\nu})$, in the range of $\omega$ close to 1. In the relevant range of parameters, the resonant contribution as function of $\omega$ takes the shape of a peak, which we call the *resonant peak*. A reasonable way to characterize the *performance* $P$ of the resonant drift for a given set of the parameters is to do this in terms of the ratio between the maximum of the resonant peak and the maximum of the Esaki-Tsu peak $\tilde{v}_{ET}(1) = 0.5$:

$$P \equiv \frac{\max_{\omega \approx 1} \{\tilde{v}_{res}(\omega)\}}{0.5}, \qquad \tilde{v}_{res}(\omega) \equiv \tilde{v}_d(\omega) - \tilde{v}_{ET}(\omega/\tilde{\nu}). \tag{S.19}$$

Our theory approximates $P$ as

$$P = 2A(\alpha), \tag{S.20}$$

where $A(\alpha)$ is given in Eq. (S.6) and presented graphically in Fig. 4(a). Our theory therefore suggests that the best performance $P_{\max}$ corresponds to $\alpha = \alpha_{\max} \approx 1.16$ and is equal to $2A_{\max} \approx 0.76$. It is therefore convenient to introduce the *performance index*

$$i_p \equiv \frac{P}{P_{\max}} = \frac{A(\alpha)}{A_{\max}}, \tag{S.21}$$

which immediately characterizes the extent of the performance (the latter equality in (S.21) corresponds to the approximation of $i_p$ by our theory).

Thus, the most important characteristic of the resonant drift is predicted by means of our rather simple *analytic* formula[††], in contrast to former approaches which required time-consuming numerical calculations for any given set of parameters $\epsilon$ and $\tilde{\nu}$.

It may also be of interest to know the parameter $\tilde{\nu}$: the smaller $\tilde{\nu}$ is, the better the separation between the ET peak and the resonant peak is. Similar to $\alpha$, the parameter $\tilde{\nu}$ is expressed explicitly via the physical parameters: see Eq. (S.18).

Like the maximum of the resonant peak, its doubly-scaled shape is entirely determined by $\alpha$: see Eqs. (S.13), (S.14) and (S.9). In the limiting cases $\alpha \ll 1$ and $\alpha \gg 1$, it reduces to a Lorentzian (see (10) or (S.10)) and a cusp-like K-shape (S.11) respectively. The half-width-at-half-maximum is given in terms of $\tilde{\nu}$ by $\Delta_{1/2}$ (S.14) and shown graphically in Fig.

---

[††]Moreover, being a universal function, $A(\alpha)$ need only be calculated once: the graphical representation of this universal function is presented in Fig. 4(a).

S.6 (its simple explicit approximations are given in Eqs. (S.15) and (S.16)). In terms of the cyclotron frequency, the half-width is a product of $\Delta_{1/2}$ and $\tilde{\nu}$.

We stress also that, prior to the present work, the solution of the inverse problem – i.e. a prediction of the ranges of physical parameters which would provide required characteristics of the drift – was very difficult even on the assumption of the validity of the semiclassical model (1): the search had to be done by "trial and error" using, for each set of parameters, a purely numerical solution of the equations of motion (5) and (7) with a further numerical calculation of the integral (6). In contrast, we can now solve the inverse problem in the much easier way that is demonstrated below. In addition, our results promise to help in the optimal choice of the parameter ranges where the model is valid, as described below.

Apart from the maximum and the width of the resonant peak, it may be of interest to control other characteristics of the peak, e.g. its "sharpness" – which in the most interesting range $\alpha \lesssim \alpha_{\max}$ may be characterized by the ratio of the maximum to the half-width-at-half-maximum. Control over this quantity will be considered elsewhere, while here we concentrate on how to control the height of the maximum i.e. on controlling the performance (S.19) or, equivalently, the performance index (S.21)).

In subsection 6.1, we illustrate control on the assumption that the semiclassical model (1) is valid over the entire ranges of the magnetic field parameters considered. We refer to this as "dynamical control". In subsection 6.2, we summarize our analysis of model validity, demonstrating that the restrictions on validity typically play a major role, and illustrating simple ways of tailoring the SL parameters so as to minimise the effect of at least some of these restrictions as well as in lowering the optimal range of $B$.

## 6.1 Dynamical control

In experiments, the most natural way to control the transport is by adjustment of the magnetic field parameters, which is of course much easier than varying the SL parameters. We demonstrate below such a control for the case of the SL parameters exploited in[3] and in our example for Figs. 1 and 3(a). As already mentioned, the analysis will be based exclusively on results obtained within the model (1).

Consider the plane $(B - \theta)$ (Fig. S.9). Using the given parameters of the SL, we express $B$ from Eq. (S.17) as function of $\theta$ for a given value of $\alpha$,

$$B \approx 10.5\alpha \frac{\cos(\theta)}{\sin^2(\theta)}, \tag{S.22}$$

and plot this function for a variety of values of $\alpha$. The plot of $\alpha = \alpha_{\max} \approx 1.16$ is especially important: for any point on this curve, the performance is the maximal possible within our theory ($i_p = 1$).

Similarly, we express $B$ via $\theta$ from the expression for $\tilde{\nu}$ in (S.18),

$$B \approx \frac{1.52}{\tilde{\nu} \cos(\theta)}, \tag{S.23}$$

and plot this function for a range of values of $\tilde{\nu}$.

Finally, we mark the boundaries of the "dynamically allowed area" i.e. the area where the conditions (S.18) hold so that the resonant transport calculated within the model (1) may be rather significant. The conditions (S.18) are not exact but, for the sake of simplicity, we choose characteristic marginal boundaries as

$$\tilde{\nu} = 1, \qquad \epsilon = 1.2. \tag{S.24}$$

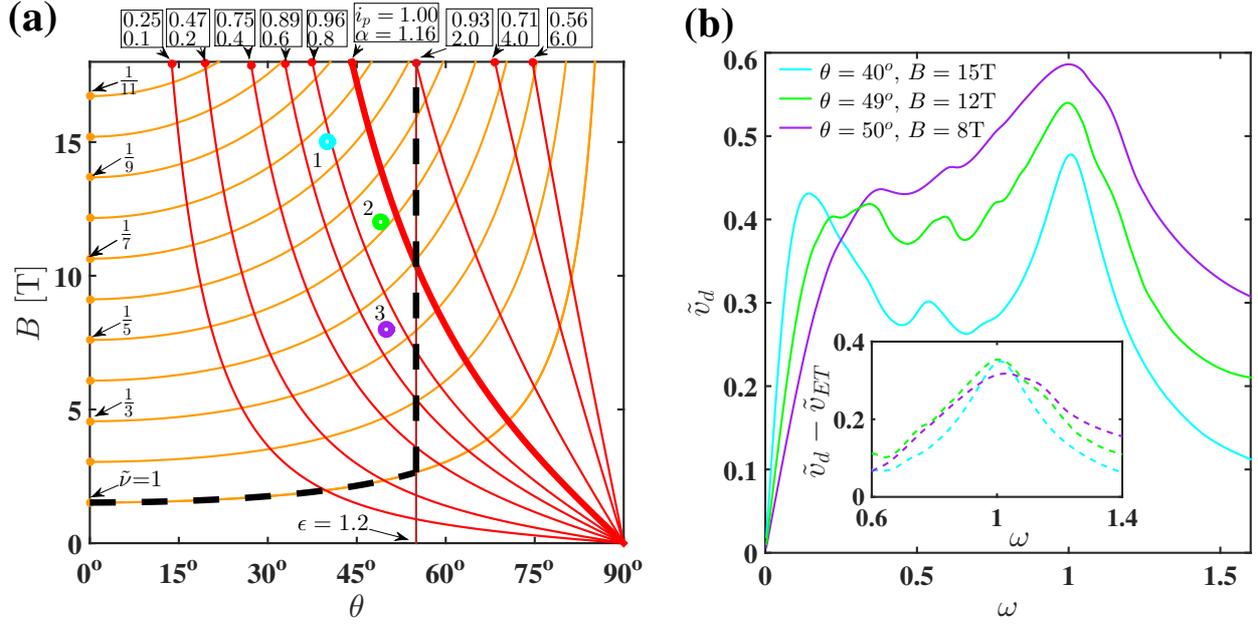

Figure S.9: (a.) The diagram in the plane of the magnetic field parameters illustrating the dynamic control over the drift (while the SL parameters are those exploited in[3] and in Figs. 1 and 3(a)). Red lines correspond to given values of $\alpha$ (indicated, together with the corresponding theoretical values of the performance index $i_p$ (S.21), near the dots where the lines meet the upper boundary of the box) and follow Eq. (S.22). The enlarged thickness marks the "best-performance" line $\alpha = \alpha_{\max}$. Orange lines correspond to given values of $\tilde{\nu}$ (indicated to the right from the dots where the lines meet the left boundary of the box) and follow Eq. (S.23). The "allowed" area, i.e. the area where our theory is at least approximately valid is limited by the line $\tilde{\nu} = 1$ from below, and by the line $\theta = 55°$ (corresponding to $\epsilon \approx 1.2$ and being marked by brown) from the right: for the sake of clarity, the boundary of the "allowed" area is marked by the black dashed line. The cyan dot indicates $(B = 15T, \theta = 40°)$ exploited in[3] and in our Fig. 3(a) while the green and purple dots indicate $(B = 12T, \theta = 49°)$ and $(B = 8T, \theta = 50°)$ respectively. The latter two dots represent examples for which the performance is approximately the same as for the green dot while $B$ is significantly smaller. (b.) The function $\tilde{v}_d(\omega)$ (cyan/green/purple solid line) calculated by (5)-(7) for $(B, \theta)$ indicated by the cyan/green/purple dot in the plate (a). The inset shows the corresponding resonant contribution $\tilde{v}_{res}(\omega) \equiv \tilde{v}_d(\omega) - \tilde{v}_{ET}(\omega/\tilde{\nu})$ (cyan/green dashed line).

The boundary conditions (S.24) are general but, for the given SL parameters, they may be presented in explicit form by using Eqs. (S.22), (S.23) and (9):

$$B \approx \frac{1.52}{\cos(\theta)}, \qquad \theta = \arctan\left(\sqrt{1.73\epsilon}\right) \approx 55°. \qquad (S.25)$$

Apart from being an immediate consequence of the conditions (S.18), the choice of the above boundaries is also based on our numerical analysis. It is illustrated in Fig. (S.10).

Fig. (S.10)(a) shows the evolution of the resonant peak for $\theta = 45°$ as $B$ increases[‡‡] which is equivalent to the decrease of $\tilde{\nu}$. At $B = 2T$, which corresponds to $\tilde{\nu} \approx 1.07$, the peak is present but its main characteristics differ significantly from those predicted by our theory: the position of the maximum is larger by about 50%, the value of the maximum is

---

[‡‡]For the sake of simplicity, we skip the range of $\omega$ to the left from the first zero: it is situated far from the relevant range $\omega \approx 1$; apart from that, the absolute value of $\tilde{v}_d - \tilde{v}_{ET}$ is small in this range.

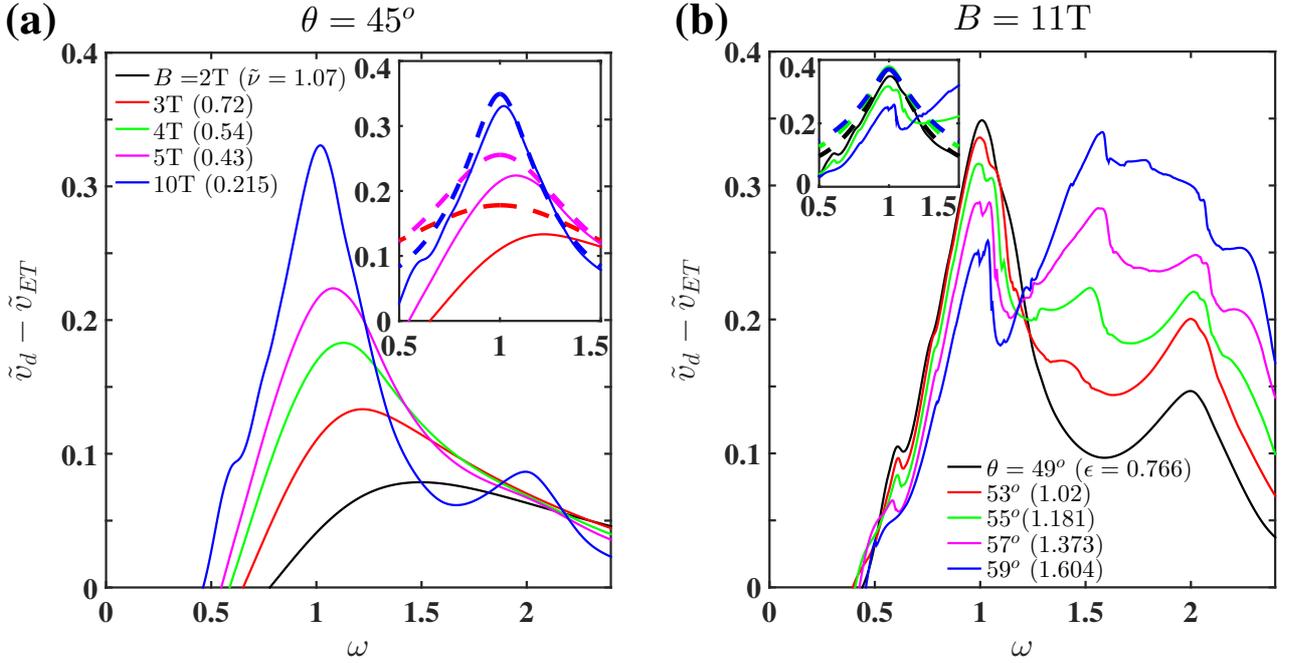

Figure S.10: Evolution of the resonant contribution $\tilde{v}_{res}(\omega)$ (S.19) as (a) $B$ increases while $\theta = 45^o$, and (b) $\theta$ increases while $B = 11T$. For the sake of simplicity, we do not show parts of $\tilde{v}_{res}(\omega)$ with $\omega < \omega_0$, where $\omega_0$ is the first zero to the left from the maximum of $\tilde{v}_{res}(\omega)$, since these parts are irrelevant. The insets compare some of the curves in the main figures (colors correspond to those in the main figures) with their approximations in our theory (dashed lines).

smaller by 40% and the width is smaller by 30%. But already at $B = 3T$, which corresponds to $\tilde{\nu} \approx 0.72$, the deviations are significantly smaller (being 20%, 23% and 23% respectively), so that the theoretical description is more reasonable. As $B$ grows further, the agreement correspondingly improves. So, Fig. (S.10)(a) demonstrates the relevance of the value $\sim 1$ as a characteristic marginal boundary in $\tilde{\nu}$ for the validity of our theory and for the presence of the distinct resonant drift.

In relation to the boundary in $\epsilon$, the situation is more subtle. It is shown in the Letter and in Sec. 5 above that, as $\epsilon$ increases and approaches the range $\sim 1$, chaos comes into play and destroys the resonant drift. However, in order for the lower boundary in $\epsilon$ with such a mechanism really to be $\approx 1$, the value of $\tilde{\nu}$ should be very small: $\lesssim 0.02$. Then chaos is already pronounced at the scattering time-scale $\nu^{-1}$ (see the example illustrated in Figs. 3(b) and 4(b) and the right-hand panels in Fig. S.8). The most interesting range of $(B, \theta)$ is that where $\alpha \sim \alpha_{\max}$ while $B$ is not too large (so that the physical model is still valid: see next subsection). In order for $\tilde{\nu}$ to be $\lesssim 0.02$ in this range of $(B, \theta)$, the scattering rate $\nu$ should be an order of magnitude smaller than that used in the experimental SLs[6,8,9] and their theoretical analogue[3]. In the latter cases, for chaos to become pronounced on the relevant time-scales, the value of $\epsilon$ should be at least a few times larger than unity. In other words, if the relevant values of $\tilde{\nu}$ are $\gtrsim 0.1$, the lower boundary for the values of $\epsilon$ which provide the required pronounced chaos shifts to values $\gtrsim 2 - 3$. However, even before that, starting from values of $\epsilon$ about $1.2 - 1.5$, the resonant peak at $\omega \approx 1$ becomes significantly smaller and, yet more important, resonant peaks at multiple and rational frequencies (initially at $\omega = 2$ and $\omega = 3/2$) grow fast and become dominant at $\epsilon = 1.2 - 1.5$ [§§]. This is illustrated in Fig. S.10(b) showing the evolution of the resonant contribution vs. $\omega$ as $\theta$ increases while $B = 11T$. The mechanism for the growth of the peaks at multiple and rational frequencies is apparently non-chaotic and to uncover it is a challenging problem for the future. In the context of our present work, it is sufficient to conclude from these numerical observations that, for the SL parameters lying in the ranges similar to those used in [6,8,9,3], the characteristic marginal boundary in $\epsilon$ may be chosen in the range $1.2 - 1.5$. We choose it to be $1.2$.

Fig. S.9(a) presents the resultant diagram of the dynamical control. There is a wide layer where the performance is high. The point $(B = 15T, \theta = 40^o)$ (marked by the cyan dot), exploited in [3] and in Fig. 3(a), does belong to this layer. At the same time, the diagram shows that $B$ may be significantly smaller while the performance still remains high. We choose two examples: $(B = 12T, \theta = 49^o)$ (marked by the green dot) and $(B = 8T, \theta = 50^o)$ (purple dot). The straightforward comparison of the resonant contributions for these three pairs of the parameters $(B, \theta)$ (Fig. S.9(b)) confirms that, despite $B$ in the green dot and, yet more so, in the purple one being significantly lower, the performance is approximately the same (even slightly higher, as for the green dot).

In conclusion, by plotting a variety of lines representing the simple explicit formulae (S.22) and (S.23), we have constructed a diagram in which we can choose a point with any desired $\alpha$ and $\tilde{\nu}$ (while the former is immediately related to the performance $P$ by the analytic formula (S.6) or, equivalently, by its graphic representation in Fig. 4(a)) within the allowed area limited by the lines $\tilde{\nu} = 1$ and $\theta = 55^o$. This allows us to easily control the resonant drift provided the physical model is valid.

## 6.2 Model validity and tailoring the SL parameters

We have carried out an extensive analysis of the validity of the model (1), based on the general literature [4,16], on the literature related to semiconductor superlattices [17,18,19,20,21], on the more specialised literature related to the particular phenomena of interest [2,6,7,22,23], and on our own calculations. The range of relevant issues includes: finite-size effects (in particular, the Hall effect), miniband conduction vs. other types of transport, inter-miniband tunneling, emission of optical phonons, quantization of motion by the magnetic field, and the charge accumulation along the SL.

A detailed account of the analysis goes far beyond the scope of this Letter and Supplementary Material (and it will be presented elsewhere). Here, we just summarize the analysis, concentrating mainly on those items which are most relevant to the experiments [6,8,9] and their theoretical analogues (e.g. [3] and some examples in our present Letter). Based on this analysis, we add into the diagram shown in Fig. S.9(a) the characteristic boundaries delineating the applicability of the model. Finally, our formulae suggest that minor modifications of the SL parameters may reduce the optimal range of $B$, diminish some of the restrictions on model validity, and possibly enlarge the area of model validity. We illustrate this through an example.

Let us first very briefly discuss restrictions which are unimportant for the experiments [6,8,9] but may be important when parameters lie in significantly different ranges. In particular, this concerns the finite-size effects – both those related to the direction along the SL axis and those related to the transverse plane. The latter include, in particular, the Hall effect. In brief, its negligibility is due to the fact that the transverse dimensions of the samples measured in[6,8,9] greatly exceeded their lengths along the SL axis (by a factor of more than 100×). Another potential restriction, shown[6,2] to be negligible for particular experiments, is inter-miniband tunneling; but of course it may be important in other cases (e.g. like those studied in[22,2]). Finally, the resonant effects for the collective electron dynamics (which determine such measured quantities as e.g. the differential conductivity) may be smeared because of the non-uniformity of the electric field along the SL, caused in turn by an accumulation of electrons towards the collector contact of the SL[22,2]. However, it was shown[22,2] that, for the sets of parameters used in the experiments[6,8,9], this effect is negligible if the effective field corresponds to the *first-order* resonance i.e. just to the most important case $\omega_B \approx \omega_c$.

---

[§§] Obviously, when pronounced chaos comes into play (which occurs at $\epsilon \sim 3$), the latter peaks also vanish.

Let us now turn to the restrictions that are of immediate relevance for the sets of parameters discussed here, namely to: (i) the cessation of the miniband-conduction (MBC) nature of the electron transport, (ii) phonon emission, and (iii) magnetic quantization.

(i). For the case when only the electric field is present, the MBC vs. the Wannier-Stark hopping and the sequential tunneling was analysed in[17] using a general quantum-mechanical approach based on nonequilibrium Green functions. It follows from the results of this work that transport characteristics calculated within the MBC approximation are reasonably consistent with those calculated within the general approach if $\omega_B/\nu \lesssim D \equiv \Delta/(\hbar\nu)$. Taking into account that the relevant electric field corresponds to the resonance, the condition for the MBC approach validity can be formulated as

$$\tilde{\nu} \gtrsim \frac{1}{D}, \qquad D \equiv \frac{\Delta}{\hbar\nu}. \tag{S.26}$$

The parameters used in the experiment[6] satisfy (S.26) and the calculations based on the MBC approach conform with the experimental results[6].

(ii). An important restriction of the model (1) may be due to the emission of the optical phonon[4,2]. If either the energy of the band transport $\Delta(1 - \cos(p_x d/\hbar))/2$ or the energy of the transverse motion $(p_y^2 + p_z^2)/(2m^*)$ reaches the optical phonon energy $E_{op}$, a phonon is emitted while the corresponding part of the electron energy drops to zero, i.e. the drift ceases. Hence, the conditions for the neglect by the phonon emission are:

$$E_{op} > \Delta, \qquad E_{op} > E_{tr}, \tag{S.27}$$

where $E_{tr}$ is a characteristic energy of the transverse motion, i.e. energy corresponding to the value of $\rho$ being $\sim 2\min(\alpha, \alpha_{\max})$. The energy $E_{tr}$ can be presented in two forms:

$$E_{tr} = \frac{\Delta}{\epsilon}\min(\alpha^2, \alpha_{\max}^2) \equiv \tag{S.28}$$

$$\equiv \hbar\omega_c \frac{D\alpha_{\max}}{4}\min\left(\frac{\alpha}{\alpha_{\max}}, \frac{\alpha_{\max}}{\alpha}\right).$$

(iii). The final significant limitation on model validity, relevant to both the experiments[6,8,9] and the theory[3], relates to the possibility of treating the magnetic field classically. It is evidently the most complicated limitation in terms of a theoretical analysis. Generally speaking, the validity of the semiclassical model as related to the treatment of the magnetic field is highly non-trivial[4] and, to the best of our knowledge, it is still not entirely clarified. For the systems of interest, it has been discussed in[21,8,2] but, only one of the criteria considered is given in analytic form. The latter relates[20,21] to the so-called magnetic-field saturation of the miniband (MFSMB). Extending the validity criterion to include a marginal range, the criterion can be formulated as $\hbar eB/m^* \equiv \hbar\omega_c/\cos(\theta) \lesssim \Delta$. For the relevant SLs, this is equivalent to the following limitation on the value of the magnetic field: $B \lesssim 11T$. At the same time, the experimentally measured maximum of the resonant peak of the differential conductivity in the SL with almost the same parameters as in our example and ($B \approx 14T, \theta = 45°$) exhibits[8] reasonable agreement with calculations based on the semiclassical model: the discrepancy is $\approx 35\%$. One may guess from the latter figure that the agreement becomes marginal (i.e. the discrepancy is $\approx 50\%$) if $B$ is further increased by a few Tesla i.e. to reach the range $16 - 18T$. Allowing for this, we introduce an empirical multiplier $1.5$ into the above MFSMB criterion: on the one hand, this multiplier is still $\sim 1$ while, on the other hand, it accords better with experiment. So, we formulate the semi-empirical criterion as

$$B \lesssim 1.5 \frac{m^* \Delta}{e\hbar}. \tag{S.29}$$

Another limiting mechanism, relevant only to a *tilted* resonant magnetic field, relates[8] to the so called magnetic-field-induced miniband (MFIM) structure arising in the resonance case. This concept is introduced in the work[8] within a sophisticated quantum-mechanical approach developed by the authors. Its presentation in[8] is very concise but only qualitative, so that in practice it is impossible to draw from it any analytic criterion in terms of the SL and magnetic field parameters. The only distinct conclusion which we may draw from this paper in the context of our model validity analysis is that their numerical calculations based on the MFIM concept, carried out for the given SL and ($B \approx 14T, \theta = 30°$), indicate that the semiclassical model should still be adequate for such parameters of the magnetic field, and the aforementioned experiment on the differential conductivity indirectly supports this inference at least qualitatively. Moreover, although the relevant validity criterion for $B$ should doubtless depend on $\theta$, such a dependence in the most interesting range $30 - 50°$ is apparently not strong: this follows from the approximate equality of the marginal values of $B$ inferred from the above calculations for $\theta = 30°$ and from the experiment for $\theta = 45°$. Given that, for the SL with the parameters relevant to our example, the marginal values of $B$ within the range for $\theta \sim 45°$ by the MFIM criterion approximately coincide with those by the semi-empirical MFSMB criterion (S.29), we conditionally combine the MFIM criterion with the MFSMB for this particular example of an SL. It can of course serve only as a very rough estimate and, moreover, we cannot say how the validity boundary in the $B - \theta$ plane by the MFIM criterion changes as the SL parameters change.

There should be one more validity criterion to be considered, related specifically to *resonant* transport. As shown in our paper, the resonant drift within the semiclassical model crucially depends on the specific classical dynamics in the transverse plane, which may be characterized as a cyclotron rotation interacting with the semiclassical degree of freedom along the SL axis, and a characteristic energy scale for such a rotation is $E_{tr}$ (S.28). On the other hand, the cyclotron rotation is known[16] to be quantized and, in the case when a charge is acted on only by a constant magnetic field, the quantum levels of energy are $\hbar\tilde{\omega}_c(n+1/2)$ where $\tilde{\omega}_c$ is the corresponding cyclotron frequency while $n = 0, 1, 2, \ldots$. Of course, our case is more complicated since the electron is not free (apart from the magnetic field, it is affected by the SL periodic potential) but, even so, the quantity $\hbar\omega_c$ may be considered as a characteristic quantum scale of the transverse energy. A classical treatment of the transverse dynamics cannot be adequate if the classical energy scale is less than the quantum one. So, we may formulate the additional magnetic-related validity criterion as follows[¶¶]:

$$E_{tr} \gtrsim \hbar\omega_c. \tag{S.30}$$

---

[¶¶]In common with the other criteria, we include the range of the marginal validity too.

We emphasize however that the criterion (S.30) does not necessarily imply that transport ceases if the criterion is violated. Rather, it marks only the validity of the model (1), where the transverse dynamics is treated quasi-classically. As for the resonant transport beyond the range of parameters determined by (S.30), it may still exist owing to other mechanisms (cf. the previous item). The latter is indirectly confirmed by the experimental results of the work[6]: a distinct resonance peak in the differential conductivity still exists for angles significantly lower than those determined by (S.30); on the other hand, the experimental and theoretical peaks differ more strongly at such angles than for angles within the range (S.30).

Returning to the condition (S.30) and using the second form of $E_{tr}$ (S.28), we can reformulate it as

$$\chi \lesssim \frac{\alpha}{\alpha_{\max}} \lesssim \chi^{-1}, \qquad \chi \equiv \frac{4}{D\alpha_{\max}}. \tag{S.31}$$

Note that the ratio $E_{tr}/(\hbar\omega_c)$ achieves its maximum value $\chi^{-1}$ just at the "dynamical best-performance" line $\alpha = \alpha_{\max}$.

In terms of $B$, the criterion (S.31) is equivalent to

$$\eta\chi \ll B \ll \eta\chi^{-1}, \qquad \eta \equiv \frac{8\hbar^2\alpha_{\max}\nu\cos(\theta)}{\Delta d^2 e \sin^2(\theta)}. \tag{S.32}$$

Thus, paradoxically, $B$ is limited not only from above but also from below. In physical terms the latter can be explained as follows. If $B$ is below the range where $\alpha \sim \alpha_{\max}$, then the increase of the rotation energy ceases due to the scattering rather than because of the saturation (characteristic of the range of $B$ at which $\alpha \gtrsim \alpha_{\max}$: see the Letter). That is why, if $B$ is sufficiently low, $E_{tr}$ is proportional to $\alpha^2 \propto B^2$ thus decreasing with $B$ faster than $\hbar\omega_c \propto B$. Hence satisfaction of the criterion (S.30) is worsening as $B$ is decreasing within the relevant range.

Although the above discussion of model validity cannot pretend to rigour, the validity criteria presented should serve at least as rough guides for the optimal choice of parameters in experiments. So, the dynamical diagram may be completed by characteristic lines according to the following criteria:

$$\tilde{\nu} = \frac{1}{D}, \qquad E_{tr} = E_{op}, \qquad B = 1.5\frac{\Delta m^*}{\hbar e}, \qquad \hbar\omega_c = E_{tr}. \tag{S.33}$$

The SL parameters of Fig. S.9(a) [***] lead to the results shown in Fig. S.11(a). One can see that the model validity restrictions play a major role. In particular, the model validity at the cyan dot, which marks the parameters used in the theoretical paper[3] and in our Fig. 3(a)[†††], is marginal – therefore the performance in experiments with such parameters may be significantly worse than that calculated within the semiclassical model. The purple dot and, yet more so, the green one are situated well inside the allowed area, so that our theoretical results may be expected to be better applicable to experiments using such parameters.

The region where high performance, the sharpness of the resonance peak and minimal restrictions of the model validity are provided optimally is $(B = 11 - 13\,T, \theta = 49 - 52°)$. The range of $B$ where the resonant drift in experiments may be expected to be reasonably well manifested and described by our theory in as wide range of angles as possible is $12 - 13T$: the relevant range of $\theta$ then is $40 - 52^o$.

---

[***]The optical phonon energy $E_{op}$ is not used in the construction of Fig. S.9(a) and, moreover, it is not explicitly mentioned in[3] (to which Fig. S.9(a) relates). However, the work[3] refers to experimental SLs[2,8,9] which use GaAs as a basic material. For GaAs, $E_{op} = 36\,\text{meV}$[24,8]. For other materials used for the SLs discussed, the phonon energy is larger[24,8]. So, the relevant value is just that for GaAs i.e. $36\,\text{meV}$.

[†††]We use this set of parameters in Fig. 3(a) for illustrative purposes: to compare our analytic theory with numerical results of[3].

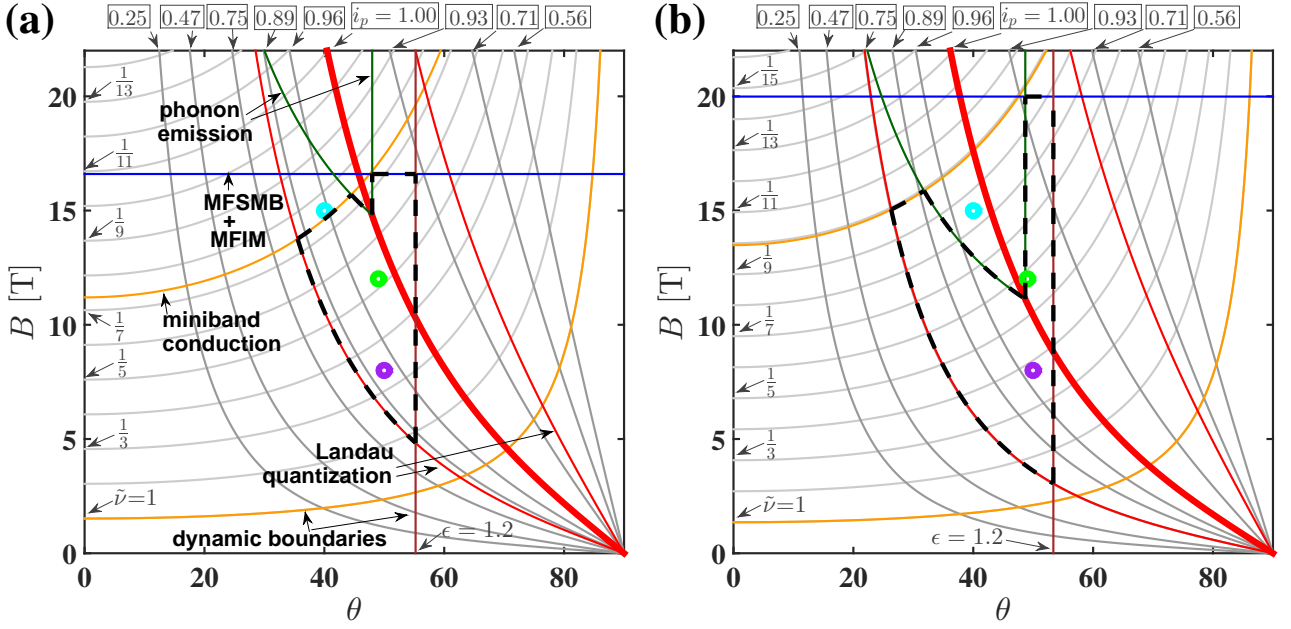

Figure S.11: (a). The diagram is analogous to that in Fig. S.9(a) but with added boundaries related to model validity: (i) the miniband conduction (S.26), (ii) the phonon emission (S.27)-(S.28), (iii) the combined semi-empirical criteria related to the magnetic-field saturation of the miniband (MFSMB) (S.29) and to the magnetic-field-induced miniband (MFIM) structure, (iv) the Landau quantization (S.30) (or, equivalently, (S.31) or (S.32)). For clarity, the lines corresponding to given values of $\alpha$ (or, equivalently, $i_p$) and $\tilde{\nu}$ are drawn in grey. The resulting boundary of the allowed area is marked by the black dashed line. (b). The diagram analogous to that in panel (a) (except for lacking an MFIM line) but with the modified SL parameters: $\nu_{new} = 3.57 \times 10^{12} s^{-1} = \nu_{old}/1.12$, $\Delta_{new} = 23 meV \approx 1.2\Delta_{old}$, $d_{new} = 8.1 nm \approx 0.976 d_{old}$.

Finally, we analyse the changes of the SL parameters which may improve the model validity, lower the optimal range of $B$ and, possibly, even enlarge the square of the area in the $B - \theta$ plane where the theory is valid. We shall assume that the basic material is the same as in the former experiments[6,8,9] (i.e. GaAs). Then $E_{op}$ and $m^*$ are fixed so that only $\nu$, $\Delta$ and $d$ may be varied. Though $d$ affects $\Delta$ (namely, $\Delta \propto 1/d$) and, to some extent, even $\nu$, the latter two quantities are also affected by other (independent) physical properties of the SL and we may therefore formally choose $\nu$, $\Delta$ and $d$ as three independent parameters.

As seen from the above analysis and from Fig. S.11(a), one of the key quantities affecting the model validity is $D \equiv \Delta/(\hbar\nu)$. Its growth both lifts the boundary related to the miniband transport validity ($\tilde{\nu} = 1/D$) and moves apart the boundaries related to Landau quantization ($\alpha = 4/D$ and $\alpha = D\alpha_{max}^2/4$). Furthermore, it leads to an increase in the largest possible value of $E_{tr}/(\hbar\omega_c) = D\alpha_{max}/4$ (achieved at the best-performance line). Altogether, in order to increase $D$, it is desirable to decrease $\nu$ and to increase $\Delta$.

A decrease in $\nu$, in addition to leading to an increase in $D$, lowers $\tilde{\nu}$ for any given point $(B, \theta)$, making the resonant peak sharper i.e. more pronounced, and decreases the gradient of lines for given values of $\alpha$, in particular the best-performance line. The latter shifts the best-performance region to lower values of $B$, which is obviously favourable too.

Growth of $\Delta$, apart from leading to the increase of $D$, lifts the validity boundary related to the magnetic-field saturation of the miniband.

On the other hand, there are restrictions on the possible decrease of $\nu$ and increase of $\Delta$. For $\nu$, the main restriction seems to be technological because of the difficulties of further reducing impurities in the crystals and of diminishing other sources of scattering (e.g. the scattering at the interfaces between the different materials used in the SL): during the eight years that passed between the papers[6] and [9], $\nu$ was decreased only by $40\%$. In the calculations for Fig. S.11(b), we use the value of $\nu$ reported in [9], which is smaller than that used in [3] and in Fig. S.11(a) by the factor $1.12$. As for $\Delta$, its increase is limited by the phonon emission mechanism. Our qualitative analysis shows that the optimal increase for the given case lies within the range $10-30\%$, depending on the particular purpose. For Fig. S.11(b), we choose $\Delta = 23\,\text{meV}$, which is approximately $20\%$ larger than that exploited in Fig. S.11(a).

The last variable parameter to consider is $d$. Its increase reduces the gradient of a line for a given value of $\alpha$: this means in particular that the region of best-performance is shifted towards lower values of $B$, which is obviously favourable. At the same time, an increase of $d$ leads to a decrease of the energy gap between the relevant (lowest) miniband and the next one[‡‡‡], which may lead to such a strong enhancement of inter-miniband tunneling that the model validity fails. An estimate of the inter-miniband tunneling rate goes far beyond the scope of the present paper. So, to guarantee the absence of the significant inter-miniband tunneling, we do not use values of $d$ exceeding those used in [7,6,3]. Moreover, the increase of $\Delta$ means that the upper boundary of the relevant (lowest) miniband is lifted for the half of the increase in $\Delta$, while the lower boundary of the higher miniband moves down (in the same proportion), so that the energy gap between the minibands decreases. Given that the values of $\Delta$ and of the higher miniband width are equal approximately[7,6,2] to $20\,\text{meV}$ and $100\,\text{meV}$ respectively, their increase by $20\%$ gives rise to a decrease in the energy gap by $12\,\text{meV}$ which constitutes only about $5-6\%$ of the initial[7,6,2] value of the gap. Most likely, such a small decrease in the gap does not violate the criterion for neglect by the inter-miniband tunneling. However, in order to guarantee the validity of this neglect, one may use a value of $d$ which is slightly smaller than the original one: this should restore the original gap. Given that the original difference in energy between the centres of the minibands is $\approx 270-300\,\text{meV}$, the decrease of $d^2$ by $4-5\%$ does compensate[16] the decrease of the gap discussed above. So, we choose $d = 8.1\,nm$.

Thus, the new parameters are:

$$\nu_{new} = \frac{1}{0.28}10^{12}s^{-1} = \frac{\nu_{old}}{1.12}, \quad \text{(S.34)}$$
$$\Delta_{new} = 23\,meV \approx 1.2\Delta_{old},$$
$$d_{new} = 8.1\,nm \approx 0.976 d_{old}.$$

The resultant diagram is shown in Fig. S.11(b). The best-performance layer has shifted to lower values of $B$ (cf. the shift of the best-performance line relatively to the cyan, green and purple dots). In particular, the optimal parameters now lie in the region ($B = 7-9T$, $49-50°$), i.e. the relevant $B$ has decreased approximately $1.5$ times; similarly, the range of $B$ for which the range of relevant $\theta$ is widest has been lowered by approximately $1.5$ times, being now $8-9T$ (the corresponding range of $\theta$ is $40-50°$). Besides, if one assumes that the modified MFIM line (which is unknown to us and is therefore absent from Fig. S.11(b)) does not significantly decrease the square of the area delineated by other modified boundary lines, then the square of the allowed area has been enlarged (even though the phonon emission region has cut off a significantly larger part of the area). Finally, the largest possible ratio $E_{tr}/(\hbar\omega_c)$ has increased by $35\%$, significantly relaxing the validity restriction in the vicinity of the best-performance line.

---

[‡‡‡]The difference in energy between neighbouring levels in the isolated quantum well is proportional to[16] $1/d^2$. In the SL, the quantum wells are not fully isolated but, still, the energy distance between centres of the minibands is approximately the same as between the levels in the isolated well, being therefore $\propto 1/d^2$.